# Livestock feeding behaviour: A review on automated systems for ruminant monitoring

José O. Chelotti[a,b], Luciano S. Martinez-Rau[a,c], Mariano Ferrero[a], Leandro D. Vignolo[a], Julio R. Galli[d,f], Alejandra M. Planisich[d], H. Leonardo Rufiner[a,e], Leonardo L. Giovanini[a]

[a] Instituto de Investigación en Señales, Sistemas e Inteligencia Computacional, sinc(i), FICH-UNL/CONICET, Argentina

[b] TERRA Teaching and Research Center, University of Liège, Gembloux Agro-Bio Tech (ULiège-GxABT), 5030 Gembloux, Belgium

[c] Department of Computer and Electrical Engineering, Mid Sweden University, Sundsvall, Sweden

[d] Facultad de Ciencias Agrarias, Universidad Nacional de Rosario, Argentina

[e] Laboratorio de Cibernética, Facultad de Ingeniería, Universidad Nacional de Entre Ríos, Argentina

[f] Instituto de Investigaciones en Ciencias Agrarias de Rosario, IICAR, UNR-CONICET, Argentina

## Highlights

- A review on monitoring methodologies of ruminant feeding behaviour is presented.
- Advantages and disadvantages of the available sensing methodologies are discussed.
- Features of the acquisition, management, and availability of the data are discussed.
- Analysis of the signal processing and machine learning methods used in the algorithm.
- Challenges and future research directions in the area are discussed.

## Abstract

Livestock feeding behaviour is an influential research area for those involved in animal husbandry and agriculture. In recent years, there has been a growing interest in automated systems for monitoring the behaviour of ruminants. Despite the developments accomplished in the last decade, there is still much to do and learn about the methods for measuring and analysing livestock feeding behaviour. Automated monitoring systems mainly use motion, acoustic, and image sensors to collect animal behavioural data. The performance evaluation of existing methods is a complex task and direct comparisons between studies are difficult. Several factors prevent a direct comparison, starting from the diversity of data and performance metrics used in the experiments. To the best of our knowledge, this work represents the first tutorial-style review on the analysis of the feeding behaviour of ruminants, emphasising the relationship between sensing methodologies, signal processing, and computational intelligence methods. It assesses the main sensing methodologies (i.e. based on movement, sound, images/videos, and pressure) and the main techniques to measure and analyse the signals associated with feeding behaviour, evaluating their use in different settings and situations. It also highlights the potentiality of automated monitoring systems to provide valuable information that improves our understanding of livestock feeding behaviour. The relevance of these systems is increasingly important due to their impact on production systems and research. Finally, the paper closes by discussing future challenges and opportunities in livestock feeding behaviour monitoring.

## 1. Introduction

Global livestock farming presents a dynamic and complex challenge. In recent decades, it has adapted in accordance with the evolving demand for animal products. Therefore, animal production systems need to increase their efficiency and environmental sustainability. The effective action in the different livestock systems depends on the advances of science and technology, which allows for increasing the number of animals caring for their health and well-being. As a result, precision livestock technologies are becoming increasingly common in modern agriculture to help farmers optimise livestock production and minimise waste and costs.

Precision livestock farming (PLF) monitors animal behaviour and disease detection at an individual level. PLF is useful to optimise animal growth and milk production by developing technologies that allow the early recognition of pathological and management-relevant behavioural changes and the assessment of the individual health state in dairy cows (Michie et al., 2020). It is a build-up of sensors, communication protocols, signal processing, computational intelligence algorithms, and embedded processors that allow the development of portable devices for real-time monitoring of individual animals, providing active management support to farming systems.

Many PLF technologies are dedicated to the study and monitoring of animal feeding behaviour. Chewing activity is a meaningful parameter of dairy nutrition to assess the adequate composition of a diet and the risk of ruminal acidosis (Yang & Beauchemin, 2007). Furthermore, the ruminating activity provides meaningful information on calving moments and subclinical diseases or health disorders (Soriani, Trevisi, Calamari, 2012). Thus, the continuous measurement of feeding variables provides a complete understanding of dietary effects on digestive function and animal performance (Dado & Allen, 1993). The timeline and intensity of feeding activity offer information on the diurnal pattern of the behaviour of ruminants, and the identification of deviations may detect health impairments (Braun, Tschoner, Hässig, 2014).

Long-term analysis of animal feeding behaviour distinguishes two main activities: rumination and grazing. These activities last from a few minutes to hours, occupying 60-80% of the daily allocation (Kilgour 2012). Their real-time account is essential for a comprehensive assessment of grazing strategies, accurate estimation of daily intakes, and detection of disease, estrus, and parturition, among other concerns. A thorough description of jaw movements (JM), the fundamental components of rumination and grazing, is crucial to achieving these objectives.

On the other hand, the design of devices for monitoring animal feeding behaviour requires a delicate balance between data acquisition, battery endurance, communication, processing, and storage capabilities. These technical requirements are related to the data to be produced and communicated (type, amount, and accuracy). Sensors allow gathering data for tracking, detecting, and classifying animal behaviours. They are usually combined with signal processing, machine learning (ML), and artificial intelligence (AI) algorithms to improve the performance of automatic feeding behaviour recognition and classification systems.

Monitoring animal feeding and locomotion activities has been done using noseband sensors (Nydegger et al., 2010; Zehner, Umstätter, Niederhauser, Schick, 2017; Werner et al., 2018), multidimensional accelerometers (Smith et al., 2016; Andriamandroso et al., 2017; Greenwood et al., 2017), inertial measurement units (IMU) and GPS (Andriamandroso, Bindelle, Mercatoris, Lebeau, 2016) and microphones (Laca, Ungar, Seligman, Ramey, Demment, 1992; Galli, Cangiano, Milone, Laca, 2011; Galli et al., 2018). It aims to alert farmers about animal behavioural changes associated with diseases, estrus, or parturition. For example, sound sensors are employed for monitoring feeding activities. They characterise the JM associated with feeding activities (Millone et al., 2011; Chelotti et al., 2016; Martinez-Rau et al., 2022), and the grazing and rumination episodes (Vanrell et al., 2018; Chelotti et al., 2020; Rau, Chelotti, Vanrell, Giovanini, 2020). Moreover, feed intake is estimated using sound energy (Laca & WallisDeVries, 2000; Galli et al., 2018; Lorenzón, 2022).

| | |
|---|---|
| AI | Artificial Intelligence |
| AE | Auto-Encoder |
| CNN | Convolutional Neural Network |
| CV | Cross-Validation |
| DA | Discriminant Analysis |
| DL | Deep Learning |
| DT | Decision Tree |
| DMI | Dry Matter Intake |
| GPS | Global Positioning System |
| GRU | Gated Recurrent Units |
| AdaBoost | Adaptive Boosting |
| ANFIS | Adaptive Neuro Fuzzy Inference System |
| ANN | Artificial Neural Network |
| BiFPN | Bidirectional Feature Pyramid Network |
| BiGRU | Bidirectional Gated Recurrent Units |
| CART | Classification and Regression Tree |
| CDA | Canonical Discriminant Analysis |
| CLSTM | Convolutional Long Short Term Memory |
| CRF | Conditional Random Field |
| ELM | Extreme Learning Machine |
| ETR | Extra Trees Regressor |
| FCM | Fuzzy C Means |
| FR | Fuzzy Rules |
| GB | Gradient Boosting |
| GBDT | Gradient-Boosting Decision Tree |
| HGBDT | Histogram-based Gradient Boosting Classification Tree |
| HMM | Hidden Markov Model |
| IMU | Inertial Measurement Unit |
| IoT | Internet of Things |
| JM | Jaw Movements |
| k-NN | k-Nearest Neighbour |
| LDA | Linear Discriminant Analysis |
| LPC | Linear Prediction Coefficient |
| LR | Linear Regression |
| LSTM | Long Short Term Memory |
| LSVR | Linear Support Vector Regressor |
| LVQ | Learning Vector Quantization |
| ML | Machine Learning |
| MFCC | Mel-Frequency Cepstral Coefficient |
| MLP | Multilayer Perceptron |
| MLR | Multinomial Logistic Regression |

| MST | Mean Shift Tracking |
|---|---|
| NB | Naïve Bayes |
| NuSVR | Nu Support Vector Regressor |
| PCA | Principal Component Analysis |
| PLF | Precision Livestock Farming |
| PLS | Partial Least Square regression |
| PLS-DA | Partial Least Squares-Discriminant Analysis |
| PNN | Probabilistic Neural Network |
| PPCA | Probabilistic Principal Component Analysis |
| QDA | Quadratic Discriminant Analysis |
| R-CNN | Region-based Convolutional Neural Network |
| RF | Random Forest |
| RFID | Radio Frequency Identification |
| Ridge | L2 regularised linear regression |
| RNN | Recurrent Neural Network |
| RSE | Random Subspace Ensemble |
| SNR | Signal-to-Noise Ratio |
| SOM | Self-organising Map |
| SR | Stepwise Regression |
| STC | Spatio-Temporal Context |
| SVM | Support Vector Machine |
| SVR | Support Vector Regressor |
| ToF | Time-of-Flight |
| TSN | Temporal Segment Network |
| XGB | eXtreme Gradient Boosting |
| YOLO | You-Only-Look-Once |

Table 1: Glossary of acronyms used in this review.

Recent advancements in hardware and image-processing algorithms have stimulated the use of videos as a monitoring technique. Fixed video cameras allow the monitoring of individual or group behaviour automatically, continuously, and non-intrusively in a given fixed area (Fuentes, Yoon, Park, Park, 2020). Their use is limited to small farm areas, such as pens and barns. On the other hand, small wearable video cameras on animals would expand the region of action, although their application still needs further development (Saitoh & Kato, 2021).

This article reviews and analyses recent trends and advances in monitoring, automatic analysis, and prediction of ruminant feeding behaviour based on different sensors/signals using a combination of signal processing and ML techniques. Articles from 2005 to 2022 were analysed using ScienceDirect and Google Scholar databases. Keywords like *machine learning*, *deep learning*, *acoustic monitoring*, *ruminant feeding behaviour*, *dairy cows*, *inertial unit*, *accelerometer*, and *precision livestock management* were employed combined to search them. These papers included related studies from science and engineering conferences, journal articles, review articles, books, theses, and other electronic document repositories. To simplify the wording of the text, numerous abbreviations and acronyms were used in this review (Table 1).

The selection criteria for the state-of-the-art techniques included the initial selection of hundreds of research articles published in the forenamed search engines. Subsequently, the selection criteria were improved by reading full-text articles to finally pick 131 articles that best fit the objective of this paper. It excludes articles based on manual techniques or direct human supervision since the latter work reported dated to 2006. The articles that analyse behaviours like reproduction or physical activities, those whose performance metrics were unavailable, or those written in languages different from English were excluded. Finally, commercial devices that have been significant for the subject were included. The technical information provided by the development teams limited the analysis.

Over fifty surveys and reviews about using ML and the Internet of Things (IoT) for PLF have been published in the last decade. The subjects of these works are diverse and cover different aspects of livestock production like welfare assessment (Chapa, Maschat, Iwersen, Baumgartner, Drillich, 2020; Spigarelli, Zuliani, Battini, Mattiello, Bovolenta, 2020; Azarpajouh, Calderón Díaz, Bueso Quan, Taheri, 2021), health monitoring (Eckelkamp 2019; Karthick, Sridhar, Pankajavalli, 2020; Alfons et al., 2020; O'Leary et al., 2020; Fan, Bryant and Greer, 2022), herd management (Cockburn, 2020; Yousefi, Rafi, Al-Haddad, Azrad, 2022; Hossain et al., 2022) and commercially available technologies (Stygar et al., 2021). They also include systems implementation (Lokhorst, De Mol, Kamphuis, 2019; Kim et al., 2021; Oliveira, Pereira, Bresolin, Ferreira, Dorea, 2021; Subeesh & Mehta, 2021; Farooq, Sohail, Abid, Rasheed, 2022) and opportunities and challenges offered by PLF (Bailey, Trotter, Tobin, Thomas, 2021; Niloofar et al., 2021; Aquilani, Confessore, Bozzi, Sirtori, Pugliese, 2022; Morrone, Dimauro, Gambella, Cappai, 2022).

Few articles introduce a general overview of PLF (Cockburn, 2020; Garcia, Aguilar, Toro, Pinto, Rodriguez, 2020; Aquilani, Confessore, Bozzi, Sirtori, Pugliese, 2022; Tzanidakis, Tzamaloukas, Simitzis, Panagakis, 2023), including relevant management topics, like animal identification, posture monitoring, body weight estimation, and estrus detection using different sensing technologies. Additional studies explored the use of wearable sensors (Lee & Seo, 2021) or motion sensors (Kleanthous, Hussain, Khan, Sneddon, Liatsis, 2022; Riaboff et al., 2022; da Silva Santos, de Medeiros, Gonçalves, 2023) for monitoring different behaviours, including feeding patterns. Wurtz et al. (2019) reviewed the papers based on machine vision technology for monitoring indoor-housed farm animals. Mahmud, Zahid, Das, Muzammil, and Khan (2021) discussed algorithms based on images/videos and deep learning (DL) methods. In this context, Andriamandroso, Bindelle, Mercatoris, and Lebeau (2016) analysed algorithms employing various sensing methods to monitor feeding behaviours and their associated parameters.

This work has deviated from the meta-analytical framework most frequently used in systematic reviews. It represents a self-contained overview based on the author's expertise and a selective review of the relevant literature in precision livestock farming. This approach allowed for a thorough examination of the topic through targeted searches of resources such as ScienceDirect and Google Scholar. This article provides three main contributions. Firstly, it introduces a detailed description of the forage intake mechanism to understand the feeding phenomenon and the advantages and drawbacks of the sensing methods employed for monitoring. This fact allows a better analysis of the advantages and disadvantages of sensing techniques. Secondly, non-invasive monitoring methodologies are analysed and compared, highlighting the advantages and disadvantages of the most ubiquitous sensors. Thus, we will focus our analysis on algorithms that provide the most relevant information about ruminants' feeding behaviour. This choice leaves out of the scope methodologies that measure internal body variables like rumen pH, temperature, and movements (Hajnal, Kovács, Vakulya, 2022). Finally, taking advantage of the multidisciplinary background of the authors, a general discussion about the current state and future challenges is presented.

The paper is structured as follows. Section 2 introduces the basis of the ruminant forage intake mechanism. Section 3 describes several monitoring methodologies based on different types of sensors. Section 4 introduces some commercial devices developed in this area. Finally, Sections 5 to 7 present the discussion, conclusions, and future works, respectively.

## 2.    The forage intake mechanism

Voluntary forage intake is one of the factors that best explains cow milk production. Cows dedicate 5 to 9 hours to grazing (spread over 10 to 15 bouts) and a similar amount of time to rumination during the day. For this reason, feeding monitoring is so relevant for the productive management of a livestock system. The number of total chews per food unit (mainly during rumination) is associated with particle size reduction and the amount of produced saliva. In this way, the nutrients available in the food are better assimilated and help to maintain an adequate rumen environment (De Boever, Andries, De Brabander, Cottyn, Buysse, 1990). These factors improve the productivity and health of the animals. Thus, changes in the daily pattern of these activities can explain the productive results and expose limiting conditions in animal production systems.

The choice of variables used to monitor and diagnose the foraging behaviour depends on the spatio-temporal integration model used as a reference. Bailey et al. (1996) proposed a conceptual model of ingestive behaviour based on six increasing levels: from the bite, the feeding station, the patch, the feeding site, the field or pasture, and up the habitat (Fig. 1). The model was modified to employ it in this work: grazing is a process that combines different movements and activities at different scales of time and space (Fig. 2.a). At the level of production systems with a certain intensification, it would be enough to integrate the scales from bite to pasture level, combining the intermediate feeding scales, to adequately describe daily forage intake for one or more days.

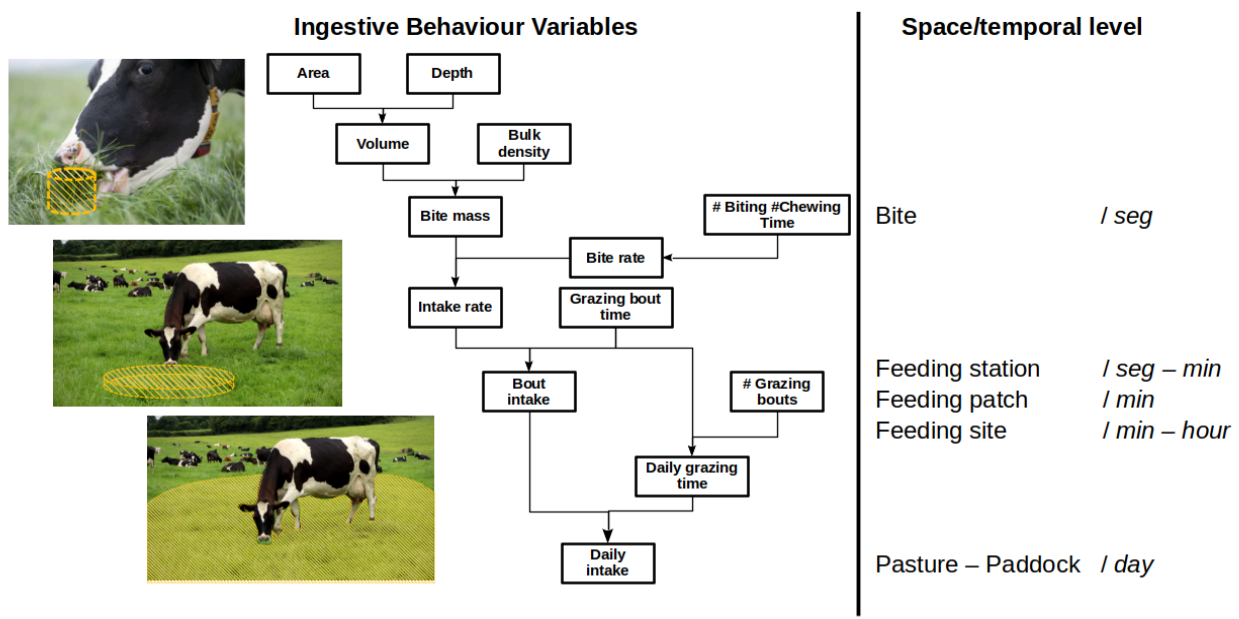

**Figure 1:** Conceptual model of ingestive behaviour and its spatio-temporal levels (adapted from Bailey et al. (1996)).

Underlying relationships between plants and animals during grazing explain the behaviour variations over time and space, which is critical for managing grasslands and pastures. The essential component of ingestive behaviour in grazing cattle is the bite. It includes the movements of apprehension and severing of forage, affected by different characteristics of the mouth (size and mass of jaws, muscle characteristics, etc.) and pasture, such as structure, leaves distribution, chemical composition (water or fibre content), and the amount of forage harvested in each bite.

Grazing at a *bite* level (Fig. 2.c) comprises three phases. Firstly, the animal approaches the pasture and sweeps around with the tongue to bring herbage into the mouth (bite apprehension). Then, it presses the forage between the lower incisors and the upper dental pad (bite cutting). Finally, it finishes harvesting each bite, tugging and breaking the forage with a quick head movement. Once a bite process concludes, the forage in the mouth is comminuted using premolars and molars in a chewing process known as *grazing chew* (Fig. 2.c). Animals execute these activities through JM (opening and closing their jaws). Each JM is associated with specific feeding actions: biting, chewing, or a compound movement that includes chewing and biting when the animal closes its jaw called *chew-bite* (Laca & Wallis DeVries, 2000; Ungar et al., 2006) (Fig. 2.c). The forage consumption process concludes when, after severing one or several bites, the chewed forage in the mouth forms a cud that generates a stimulus to swallow it.

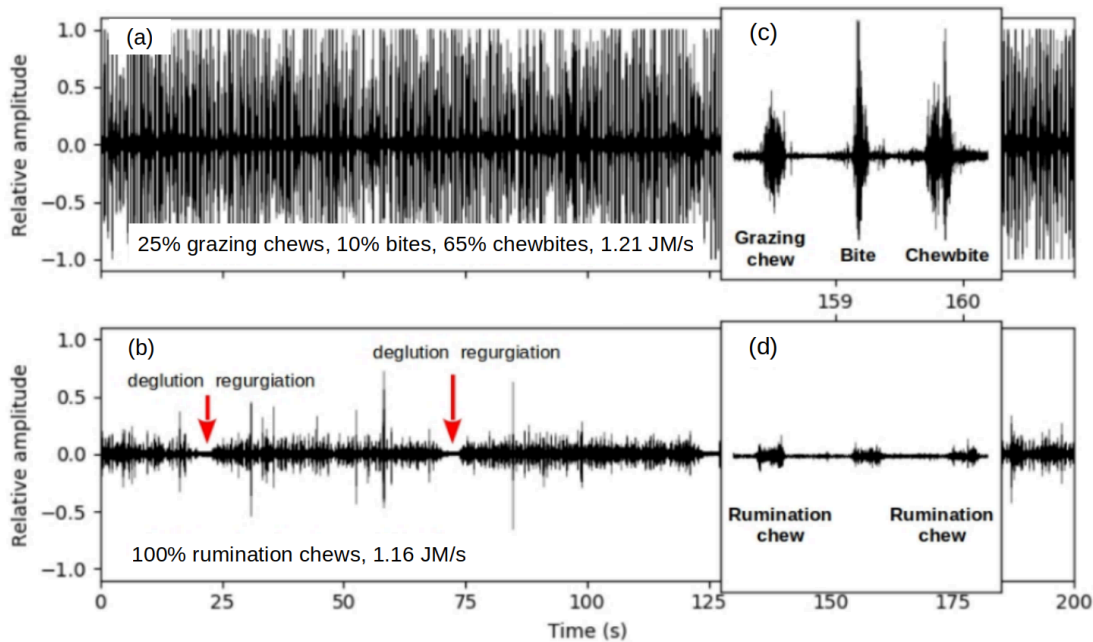

**Figure 2:** Sound recorded during (a) grazing and (b) rumination activities, including representative JM ratios and rate (JM s$^{-1}$) by activity (adapted from Chelotti et al., 2020).

The bite volume, defined by the bite area and depth, and the forage density determine the amount of forage reaped in each bite (Laca, Ungar, Seligman, Ramey, Demment, 1992; Ungar et al., 2006). The average bite mass (grams per bite) and the bite rate (bites per unit of time) determine the speed of animal forage ingestion or *intake rate*. Finally, the daily intake will be the product of the intake rate by the effective hours that animals graze per day (daily grazing time). Daily grazing time is the accumulation of grazing bouts performed during the day (Fig. 1).

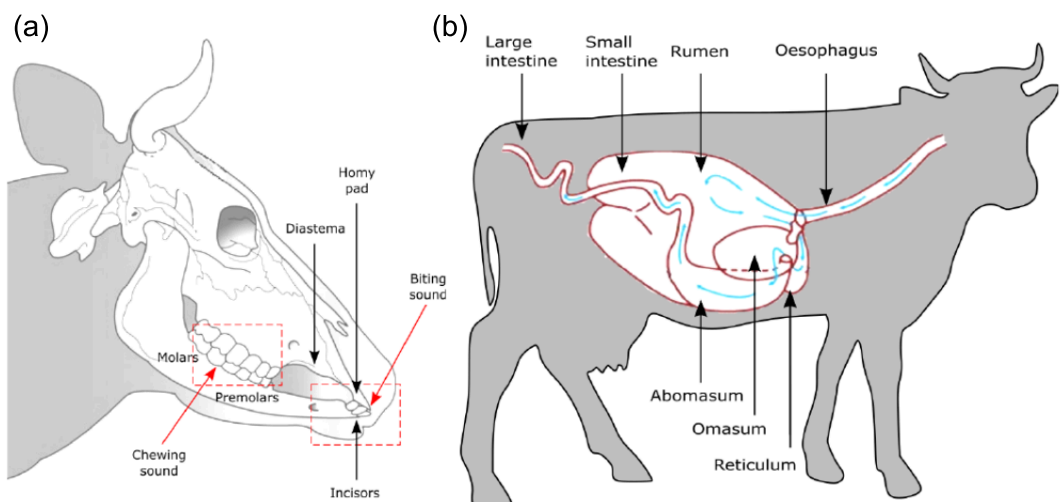

**Figure 3:** Diagrams of a) the jaw with places that produce ingestive sounds and (b) the digestive system.

Like grazing, rumination occurs in spaced regular sessions throughout the day. During rumination, ruminants no longer need to move their heads to harvest and grind herbage. Food particles are sorted in the rumen by the reticulum-rumen (Fig. 3.b) generating a bolus that is reprocessed in the mouth to decrease their size, increasing the food surface-to-volume ratio. Rumination only requires JM to crush the rumino-reticular bolus. It is composed of three phases (Fig. 2.b): regurgitation when the animal regurgitates a bolus to the mouth; jumbling and binding when the animal chews and salivates the bolus in the middle region of the jaws using molars and premolars

(Fig. 3.a); and deglutition when the animal swallows the bolus. During the second phase, the animal performs a JM known as *rumination chew*. Rumination bouts last between 45 s to 70 s, containing 30 to 60 rumination chews with a minor variation in their number. Rumination bouts are repeated uninterruptedly during a rumination session. Daily rumination time is the aggregation of all rumination sessions. The rumination process stimulates saliva secretion to help buffering the rumen pH, reduce forage particle size, and improve rumen bacteria to attach to forage particles during microbial fermentation (De Boever, Andries, De Brabander, Cottyn, Buysse, 1990).

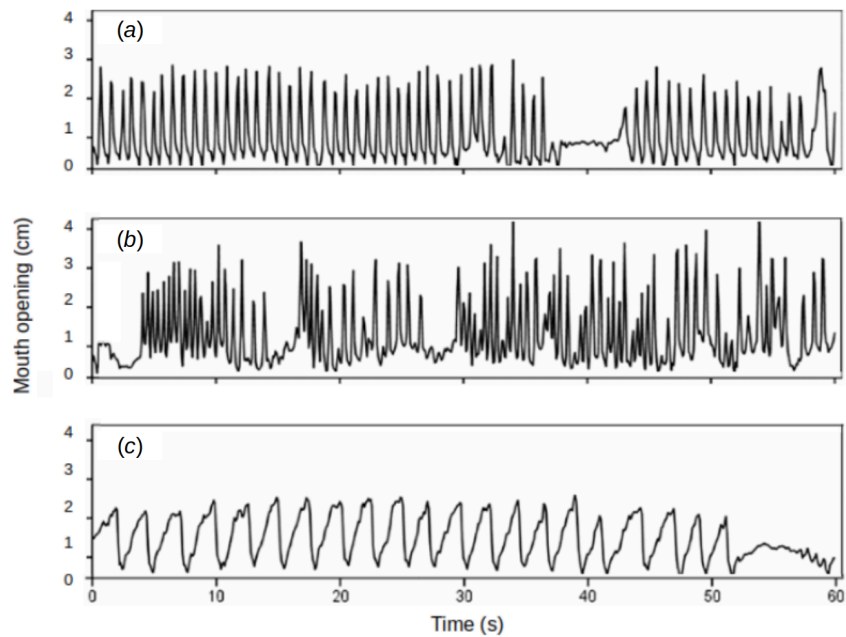

**Figure 4:** Time series of typical mouth opening for a) rumination, b) grazing, and c) drinking (adapted from Zehner, Umstätter, Niederhauser, Schick, 2017).

The biomechanical characteristics of the mouth (size and mass of jaws, muscle characteristics, etc.), the saliva and food availability, and the forage density determine the JM rate (Virot, Ma, Clanet, Jung, 2017). The mouth opens between 2 and 4 cm for rumination, grazing, and drinking (Fig. 4). The JM rate during grazing ranges from 0.75 to 1.20 JM s$^{-1}$ (an average of 1.00±0.25 JM s$^{-1}$), while it has an average of 1.06±0.06 JM s$^{-1}$ during rumination (Andriamandroso, Bindelle, Mercatoris, Lebeau, 2016). Food availability and characteristics (sward height, tensile strength, and bulk density) explain the greater JM rate variation during grazing.

JM and food characteristics (fibre content, tensile strength, water content, and density) determine the distinctive features (shape, intensity, energy, and frequency content) of sounds produced during JM. Sounds associated with grazing chews have moderate energy, moderate amplitude, and middle duration (Fig. 2.c). They arise in the middle region of the jaws (Fig. 3.a), where premolars and molars grind the forage. The rupture of the plant cells and the extrusion of internal water content determine the energy of the sound (Galli, Cangiano, Demment, Laca, 2006). Sounds associated with bites have moderate energy, high amplitude, and short duration (Fig. 2.c) because of herbage tearing and cutting. These sounds originate when the animal cuts the plants with the lower incisors and the upper horny pad (Fig. 3.a). Finally, sounds associated with chew-bites combine bite and grazing chew features, resulting in a sound of high amplitude and energy, and long duration (Fig. 2.c). In penning systems, ruminants do not need to perform all the grazing phases because forage is supplied in feeders or on the ground. They just need to chew and manipulate the food to swallow it.

Sounds associated with rumination chews have low energy, low amplitude, and middle duration (Fig. 2.d) due to the chewing of the cud. Its energy and amplitude are low because grass fibres have incorporated extra water (during their dwellings in the rumen) and have already broken down. The sounds arise in the middle region of the jaws (Fig. 3.a), and regurgitation and deglutition pauses produce very low-intensity sounds (Fig. 2.b).

# 3.    Monitoring and analysis methodologies

Ruminants perform specific body and head movements and produce distinctive sounds when grazing and ruminating. Monitoring techniques record and analyse these movements and sounds to characterise ruminants' feeding activities. Thus, monitoring techniques are classified according to the technique used to record the movements and sound:

1. **Motion:** Feeding activities are estimated indirectly by sensing body movements and postures (Brennan, Johnson, Olson, 2021, among others) and movements (Tani, Yokota, Yayota, Ohtani, 2013) through motion sensors. In other cases, JM can be directly measured by sensing changes in pressure or length of a sensor around the nose (Nydegger et al., 2010; Chen, Li, Guo, et al., 2022, among others). All these devices are wearable sensors;

2. **Sound**: JM can be characterised indirectly by recording and analysing the sound patterns produced during feeding activities (Milone, Galli, Cangiano, Rufiner, Laca, 2012; Navon, Mizrach, Hetzroni, Ungar, 2013; Chelotti et al., 2016; Chelotti et al., 2018, among others). Different types of microphones are wearable sensors; and

3. **Images**: Imaging systems sense and monitor the body movements and postures associated with feeding activities (Gu et al., 2017; Hansen, Smith, Smith, Jabbar, Forbes, 2018, among others). Cameras are employed either in fixed positions or as wearable devices.

Wearable sensors are the most widely used acquisition devices to cover large areas of farms and fields. However, operational requirements (device portability, robustness, and energy capacity) and the computational cost of algorithms typically pose challenges to further technological development and adoption (Stone, 2020). Other important considerations include the specificity of the sensor placement on the animal body and the surrounding environmental noises and disturbances that can negatively impact signal acquisition (Fig. 5).

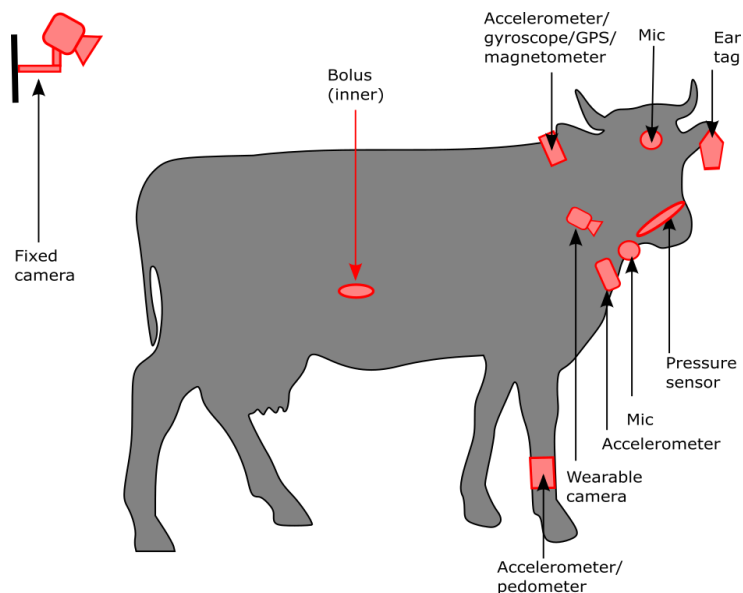

**Figure 5:** Typical placement of sensors and devices for monitoring feeding behaviour.

Several algorithms have been developed in the last decade to analyse the information provided by sensors (microphones, pressure sensors, accelerometers, cameras) used to monitor the ruminants' feeding behaviour. They are pattern recognition systems that aim at classifying input data (pressure, sound, accelerations, and images) into a set of specific classes of JM (ruminating chew, grazing chew, bite, and chew-bite) and feeding behaviours (grazing, ruminating, others).

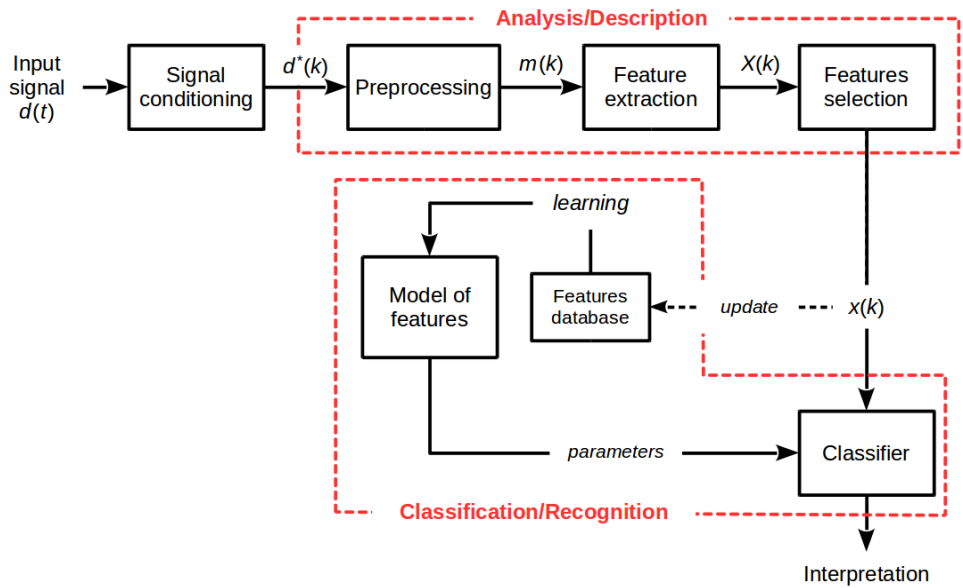

**Figure 6:** Block diagram of a general pattern recognition system.

A pattern recognition system implements a series of generic stages (Fig. 6) that allows: i) the description and analysis of the input signal through distinctive features that simplify (ii) their recognition and organisation into classes, enabling the identification of patterns (Duda, Hart, Stork, 2000). The first stage of a pattern recognition system is signal conditioning, where the input signal $d(t)$ is prepared to meet the system requirements. It uses analogue and digital signal processing techniques to transform $d(t)$ into $d*(k)$. The preprocessing stage processes $d*(k)$ to simplify the extraction of features and to reduce the computational load by transforming $d*(k)$ into the segmented signal $m(k)$. The goal of the *feature extraction* stage is to characterise events using features $X(k)$, arranging the events into classes by seeking $X(k)$ that unequivocally identified $d*(k)$ with each event. Finally, *feature selection* optimises $X(k)$ to improve and simplify the classification task by retaining the features that boost discrimination and by removing the others. This transformation of $d*(k)$ into $x(k)$ can be "continuous" (window-based) or triggered by specific events (event-based). The classifier is trained and its parameters are tuned using a portion of the database. After a successful learning process, the classifier uses $x(k)$ as input to identify patterns and then generate the output of the system, which is often organised into categories or classes.

There are two approaches for training models: learning its parameters from a training dataset assembled from a database (*offline learning*) or updating the training dataset and the parameters every time new data is available (*online learning*). Each of these approaches has advantages and drawbacks. Their applicability depends on the features' nature: time-varying features require online learning, while time-invariant ones need offline learning. Model development ends with its testing and validation (Fig. 7).

A central part of any pattern recognition system is ML. Figure 7 shows a typical ML workflow to create a model. Data collection and preparation are the first tasks in this process. Data curation is required to develop a model with good performance. The curated data are split into three independent datasets (training, testing, and validation) to be used in the following task of the development process. The candidate model and training algorithms are chosen based on the characteristics of the problem. The candidate model is trained using the training data and evaluated using the validation data. Performance metrics associated with the model and collected data distribution measure the model performance to choose the best trained model. The next task consists of evaluating the chosen model using the test data. A poorly performing model may require retraining. Contrary, a tested model providing solid performance achieves appropriate training, indicating good data generalisation capabilities. Finally, the model is deployed and sent to production. Its performance is monitored along its deployment in case it may require retraining.

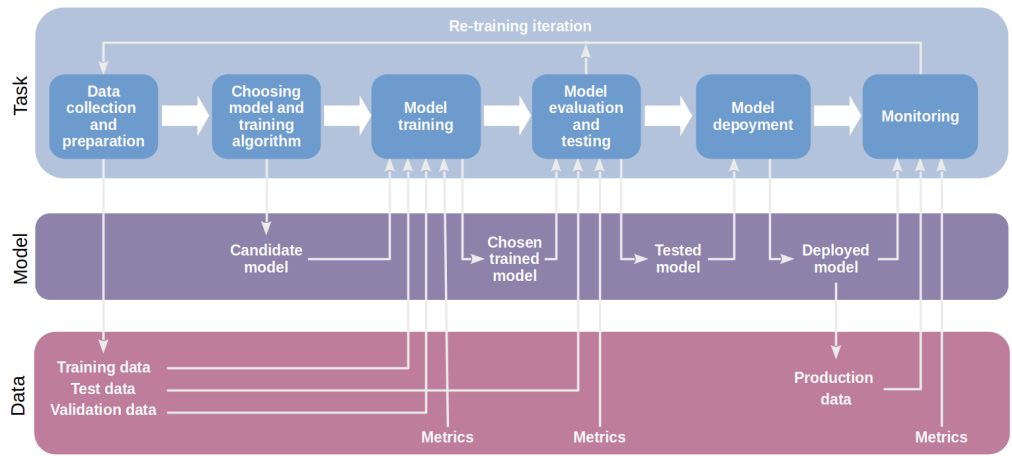

**Figure 7:** ML workflow (adapted from https://www.altexsoft.com/).

Articles found in the literature follow different approaches to develop their models. Two methodological issues, regardless of the classification/regression problem, linked to the training and the performance evaluation of ML models must be taken into account (Sokolova & Lapalme, 2009). Firstly, it must be analysed how to split the dataset for training and testing/validation (Fig. 7). A simple approach splits the dataset into two subsets: one for training and another for testing/validation. It is referred to as *holdout validation*. The model parameters are adjusted using the training dataset, while the testing/validation dataset is used to evaluate the resulting model. It usually includes a classification bias in the model since it is validated using a subset of the original dataset. The most commonly used method is the *k-fold cross-validation* (CV). This method involves dividing the dataset into *k* groups (folds), and the training-evaluation process is repeated *k* times. In the *i-th*-iteration ($1 \leq i \leq k$), the *i-th-fold* is used to test the model, while the remaining folds are for training. A third approach integrates the previous ones by initially dividing the original dataset into two parts one for training and the other for testing. The model parameters are then tuned using a *k*-fold CV approach with the training set. Finally, the testing set, independent of the training one, is used only to report the final results.

The robustness of classifiers can be enhanced by incorporating data from animals different from those used for model training. During a CV process, a common strategy involves training the model with specific animals and reserving one animal for evaluation (called *leave-one-animal-out*). Some authors suggest a similar approach but utilise data from more than one animal in each fold without grouping data from a single animal in more than one fold (Pavlovic et al., 2021). Other authors applied a similar concept but at the level of signals and independently of the animal. When the dataset is small, the *leave-one-signal-out* methodology is usually employed (Milone, Galli, Cangiano, Rufiner, Laca, 2012; Chelotti et al., 2018), utilising one signal for evaluation and the remaining signals for training.

Finally, a typical aspect of this type of problem is the skewed class proportion in the dataset, known as class imbalance (Hasib et al., 2020). This occurs when one class is much more abundant than the others. In such cases, models tend to predict the majority classes but may fail to accurately capture the minority ones. Resampling is a widely adopted technique for highly unbalanced datasets (Sakai, Oishi, Miwa, Kumagai, Hirooka, 2019; Fogarty, Swain, Cronin, Moraes, Trotter, 2020; Watanabe et al., 2021). It involves either removing samples from majority classes (*under-sampling*) or adding synthetic examples to minority ones (*over-sampling*). In these cases, the model is evaluated using metrics aimed at avoiding bias toward the majority classes, such as the area under the operation curve, characteristic curve, confusion matrix, precision, recall, and F1-score (Ali, Shamsuddin, and Ralescu, 2015).

In the following subsections, several of the above-mentioned aspects associated with the most popular sensing technologies in the field (i.e. motion sensors, sound sensors, image sensors, and others) are described in terms of a general pattern recognition system (Fig. 6).

## 3.1.    Motion sensors

Movement sensors have been extensively used to monitor livestock activities by identifying the ruminants' behaviours based on their head and body postures and movements. Ungar et al. (2005) introduced ML techniques for feeding activity recognition. Since this seminal work, many authors have employed ML techniques to estimate feeding behaviours alone (Yoshitoshi et al., 2013; Schmelinga, Elmamoozc, Nicklasc, Thurnera, Rauchb, 2021), alongside other types of behaviours (Dutta, Smith, Rawnsley, Bishop-Hurley, Hills, 2014), and in combination with others (Nielsen, 2013).

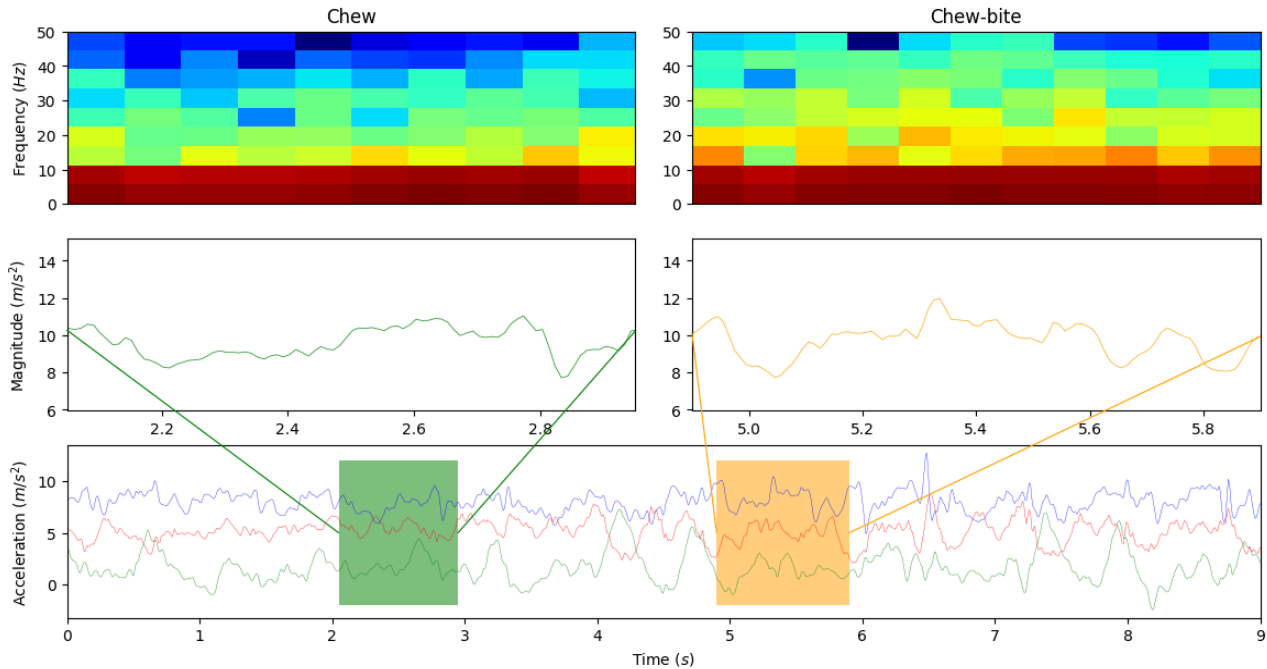

**Figure 8:** Acceleration signals recorded during a grazing period and spectrogram obtained from the magnitude vector.

Through a 3D accelerometer located on the neck, along with the magnitude and spectrograms of two individual events, it is possible to capture typical signals from grazing cattle (Fig. 8). Following the processing of these graphical records, subtle differences associated with each JM can be detected, allowing for very accurate identification and classification.

### 3.1.1.    Data acquisition and management

Feeding behaviour studies require large amounts of reliable data. Gathering them is a complex and extensive task that requires significant logistics and efforts to plan and conduct field experiments, usually under difficult environmental conditions. Due to the magnitude of this effort, a small number of authors record their particular databases and make them available online (Vázquez-Diosdado et al., 2015; Barker et al., 2018; Hamilton et al., 2019; Pavlovic et al., 2021; Li, Cheng, Cullen, 2021). Creating a database involves performing experiments, collecting data, and meticulously curating and labelling them. The labelling process requires ground-truth references. Direct visual observation is a dependable (although tedious) method to generate such references. Its complexity increases with the number of animals and the data-collecting period (Elischer, Arceo, Karcher, Siegford, 2013). Thus, researchers usually use video records to reduce mislabelling when animals are spatially confined in indoor environments (Peng et al., 2019; Shen et al., 2021) or in closed grazing patches or paddocks (Barwick, Lamb, Dobos, Welch, Trotter, 2018; Kamminga et al., 2018), where multiple fixed cameras can be employed. To simplify this task and to expand the collection period, some studies use commercial sensors to gather ground-truth references (Pavlovic et al., 2021; 2022). The quantity of data collected in the experiments depends on parameters like i) the number of animals, ii) the data collection period, and iii) the experiment duration, among others. They vary from study to study, requiring clear rules for their selection. In the papers considered in this review, it was found that the number of animals ranges from 3 (Guo, Welch, Dobos, Kwan, Wang, 2018; Hamilton et al., 2019; Li, Cheng, Cullen, 2021) to 225 (Jung et al., 2021), the collecting period ranges from 57 h (Riaboff et al., 2020) to 403 h (Hamilton et al., 2019), and the period ranges from 1 day (Roland et al., 2018) to 31 days (Gonzales, 2015). Different ruminant species were analysed in these studies.

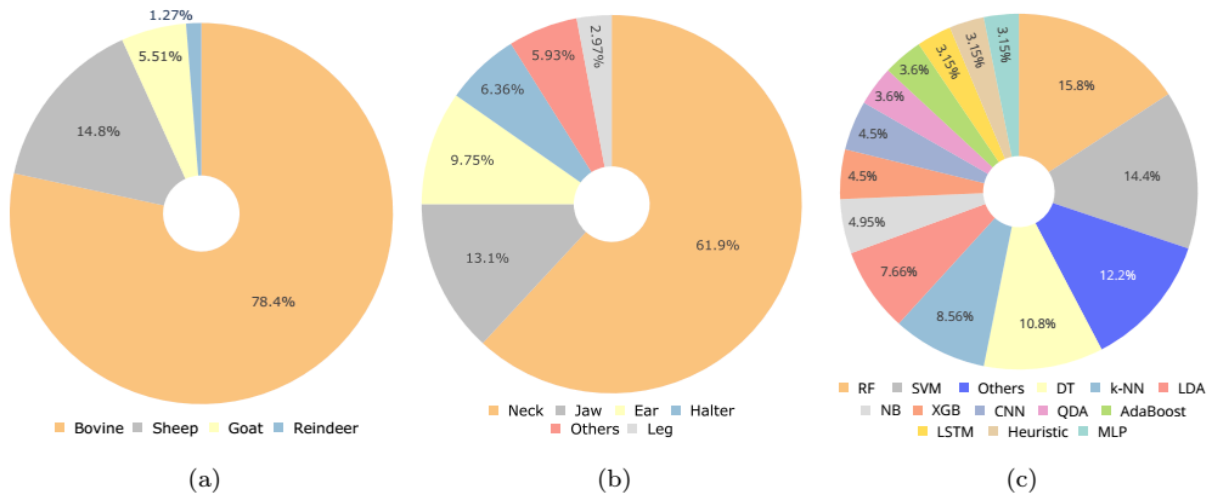

(a)  (b)  (c)

**Figure 9**: Ruminant species considered in the bibliography for movement monitoring (a). Sensor locations for feeding behaviour monitoring using motion sensors ("Others" item includes nasal bridge, horn, chest, Calan broadbent, rumen, forehead, and back) (b). ML methods used for motion-based monitoring techniques (c).

The reviewed articles show varying proportions for each ruminant species (Fig. 9.a). Bovines are the most commonly employed species in 78% of the studies, followed by sheep with almost 15%. Goats and reindeer are the less explored species, employed in 5% and 1% of the works, respectively.

The selection of the motion sensor is a fundamental aspect of activity recognition as it determines the type of information used. Initial studies employed commercial collars based on global positioning systems (GPS) (Ungar et al., 2005; Augustine & Derner, 2013) or accelerometers (Martiskainen et al., 2009; Nielsen, 2013; Yoshitoshi et al., 2013). They record head and body postures and movements. In the last decade, motion sensors based on accelerometers have been broadly adopted by researchers since they are easy to use and robust (Rayas-Amor et al., 2017; Kasfi, Hellicar, Rahman, 2016; Benaissa, Tuyttens, Plets, De Pessemier, et al., 2019; Hamilton et al., 2019; Shen et al., 2021; Pavlovic et al., 2021; 2022). Additional sensors are usually included in the devices to improve activities recognition. Accelerometers and gyroscopes located in the neck are employed to obtain supplementary information on head movements (angular velocity) as well as position (angle) (Smith et al., 2016; Andriamandroso et al., 2017; Guo, Welch, Dobos, Kwan, Wang, 2018; Mansbridge et al., 2018; Carslake, Vázquez-Diosdado, Kaler, 2020; Li et al., 2022). Furthermore, magnetometers provide information on head orientation (Kleanthous et al., 2018; Peng et al., 2019). Accelerometers and GPS are used together to track the cattle herds' locations and spatial scattering (Cabezas et al., 2022) and to improve recognition tasks (González, Bishop-Hurley, Handcock, Crossman, 2015; Brennan, Johnson, Olson, 2021). Finally, studies combined an accelerometer with either a force sensor (Decandia et al., 2018) or a temperature sensor (Dutta, Natta, Mandal, Ghosh, 2022; Fonseca, Corujo, Xavier, Gonçalves, 2022) to improve feeding activity recognition.

The sensor location determines the type of behaviours the device can identify, enabling it to identify feeding behaviours (Nielsen, 2013; Riaboff et al., 2020; Arablouei et al., 2021), diverse behaviours (Vázquez-Diosdado et al., 2015; Arcidiacono, Porto, Mancino, Cascone, 2017; Rahman et al., 2018; Roland et al., 2018; Tamura et al., 2019), or behaviours and locomotion (Martiskainen et al., 2009; Rahman et al., 2016; Alvarenga et al., 2016; Barwick, 2018; Riaboff et al., 2019; Fogarty, Swain, Cronin, Moraes, Trotter, 2020; Carslake, Vázquez-Diosdado, Kaler, 2020; Li, Cheng, Cullen, 2021). Its optimal location has been assessed in several studies (Rahman et al., 2018; Barwick, 2018; Ding, 2022). Many studies place the sensor around the neck (at its top -Arcidiacono, Porto, Mancino, Cascone, 2017- at its bottom -Bishop-Hurley et al., 2014; Brennan, Johnson, Olson, 2021- or at its side -Riaboff et al., 2019; Riaboff et al., 2020-). Other studies install the sensor either at the side of the jaw (Nielsen, 2013; Rayas-Amor et al., 2017; Shen et al., 2021) or under it (Alvarenga et al., 2016; Decandia et al., 2018; Giovanetti et al., 2017; 2020). Another common location for motion sensors is the ear, within a tag (Roland et al., 2018; Fogarty, Swain, Cronin, Moraes, Trotter, 2020; Simanungkalit et al., 2021; Chang, Fogarty, Swain, García-Guerra, Trotter, 2022). Some authors explore atypical positions such as the leg (Wang, He, Zheng, Gao, Zhao, 2018; Benaissa, Tuyttens, Plets, De Pessemier, et al., 2019; Tran, Nguyen, Khanh, Tran, 2022), the upper part of the back (Sakai, Oishi, Miwa, Kumagai, Hirooka, 2019), or the skin near the rumen (Hamilton et al., 2019). The accuracy of recognition tasks is improved if the devices use multiple sensors placed in different locations (Benaissa, Tuyttens, Plets, De Pessemier, et al., 2019; Pavlovic et al., 2021; 2022).

Figure 9.b shows the locations of the motion sensors used in the literature. The most common mounting site is the neck because it is easy to fix and provides information about head position (relative to the ground) and movements, which allows the recognition of feeding activities. The second preferred site is the lower jaw because sensors provide direct information on JM (Shen et al., 2020). However, it is difficult to mount and fix sensors in this place. Finally, the ear is the third preferred mounting location because it is easy to install and provides information about the position (relative to the ground) and the movements of the head. However, the measurements are disturbed by continuous ear movement. These three places comprise 82% of the studies.

Finally, the sensor attachment (hold and orientation) is another major issue since it can introduce errors and biases that affect the recognition task. An unsuitable subjection can lead to sensor rotations or displacements during the experiments that disturb the measurements, diminishing the performance of recognition algorithms (Li, Cheng, Cullen, 2021). Ensuring the proper sensor location and orientation during a study is a challenging task. Furthermore, techniques for orientation compensation do not guarantee good results, increasing the readability and complexity of recognition algorithms (Kamminga et al., 2018).

### 3.1.2. Preprocessing

The preprocessing stage conditions the sensor signal, generating alternative signals with more useful information, and segments it. Motion signal conditioning involves the interpolation of missing values (Martiskainen et al., 2009) and the removal of outliers (González, Bishop-Hurley, Handcock, Crossman, 2015), gravity acceleration, and biases (Rahman et al., 2016; Smith et al., 2016). The execution of these tasks depends on the quality of the recorded signals, which rely upon the experiments performing conditions (weather, environment, sensor quality, and recording device). Usually, researchers execute a priori data analysis to assess its quality and accordingly define the tools and techniques to condition the data. Then, new signals are estimated to reduce the computational load of the following tasks and to improve activity recognition. Examples of this concept include the computation of the vector magnitude (Alvarenga et al., 2016; Barker, 2018) and the magnitude area (Alvarenga et al., 2016) from three-dimensional acceleration and rotational speed measurements (Mansbridge, 2018; Benaussa, 2019).

The segmentation stage divides the new signals into fixed-length segments (*windows*) of arbitrary fixed length (Dutta et al., 2015; Martiskainen et al., 2009; Barwick, 2018). Few studies explore the effect of window length on recognition performance (Andriamandroso et al., 2017; Decandia et al., 2018). Hu et al. (2020) simultaneously use several windows of different sizes with promising results. Similarly, the accepted approach is arbitrarily fixed window overlap (Arablouei et al., 2021; Li, Cheng, Cullen, 2021; Cabezas et al., 2022), but few studies explore its effect on the system performance (Riaboff et al., 2019).

### 3.1.3. Feature extraction

The feature extraction stage computes new signals, known as *features*, from segments generated in the conditioning stage. The idea is to univocally characterise the JM or behaviour, arranging them into classes. The features are computed either in time or frequency domains (Cong Phi Khanh, Tran, Van Duong, Hong Thinh, Tran, 2020).

Frequency-domain features are estimated from the frequency representation of motion signals using the Fast Fourier Transform. Then, statistical characteristics of the frequency representation are computed (mean, standard deviation, skewness, kurtosis, maximum and minimum, energy, and entropy) as features (Rahman et al., 2016; Smith et al., 2016; Rahman et al., 2018). Some authors use spectral data like the fundamental frequency (Smith et al., 2016) and specific bands (Bishop-Hurley et al., 2014) to extract additional features.

Time-domain features are computed from raw or conditioned signal segments using statistics, signal processing, or ML (self-learned) tools. Measured signals are directly employed when data segments provide discriminative information that can be used by the classifier, like position or velocities (Nielsen, 2013; Wang, He, Zheng, Gao, Zhao, 2018). When raw data does not have enough discriminative information, statistical features of the data segment are usually computed (Martiskainen et al., 2009; Dutta, Smith, Rawnsley, Bishop-Hurley, Hills, 2014; Bishop-Hurley et al., 2014; González, Bishop-Hurley, Handcock, Crossman, 2015). The most accepted statistics are the mean, standard deviation, median, quartiles, minimum and maximum value, entropy, kurtosis, and skewness. Researchers also used time-domain features computed with signal processing methods like energy (Dutta, Smith, Rawnsley, Bishop-Hurley, Hills, 2014; Bishop-Hurley et al., 2014), zero-crossing rate (Kamminga et al., 2018), or average intensity (Barwick, 2018; Riaboff et al., 2019).

Feature analysis can be a time-demanding and complex task. Thus, many authors developed automatic feature analysis methods to simplify this task. They use auto-encoders (Rahman et al., 2016) and convolutional neural networks (CNNs) (Kaski, 2016; Peng et al., 2019; Li, Cheng, Cullen, 2021; Pavlovic et al., 2021) to process the raw data for determining the set of features to be used by the system.

Time-domain features based on statistics are the most frequently used in the literature because they are easy to compute. However, they are often supplemented with frequency-domain or self-learned features to improve recognition performances (Rahman et al., 2016; Smith et al., 2016; Kamminga et al., 2018).

### 3.1.4. Classification

The goal of the classification stage is to build and validate a model to classify the behaviour from the features obtained in the feature extraction stage. The classification model can be categorised, according to the tools employed to build it, into heuristic methods, classic ML methods, and DL approaches (Fig. 9.c). Classical techniques are the most commonly used (76%). Random Forest (RF), Support Vector Machine (SVM), Decision Tree (DT), and k-Nearest Neighbors (k-NN) are the preferred ones, comprising 51% of the published works. DL (8%) and Heuristics (3%) follow classic ML techniques in researchers' preferences. Some authors use only one of these methods (Foldager, Trénel, Munksgaard, Thomsen, 2020; Ramirez-Agudelo, Bedoya-Mazo, Posada-Ochoa, Rosero-Noguera, 2022), while others compare several methods to find the most suitable one (Eikelboom et al., 2020; Schmeling et al., 2021).

Heuristics methods discriminate JM and animal feeding behaviours using simple empirical rules and thresholds for evaluating features to perform classification (Arcidiacono, Porto, Mancino,

Cascone, 2017). They are usually assigned manually, given observational data, derived from expert knowledge, or estimated from feature distribution (Porto, Castagnolo, Mancino, Mancuso, Cascone, 2022).

Classic ML methods encompass statistical inference and ensemble models. Statistical inference methods use statistics tools to classify either motion patterns from raw data of motion (acceleration, rotation, and position) or computed features. Statistical methods include models like Linear Regression (LR) (Rayas-Amor et al., 2017; Simanungkalit et al., 2021; Ding et al. 2022), Logistic Regression (Arablouei et al., 2021), and HMM (Vázquez-Diosdado et al., 2015; Pavlovic et al., 2022; Rautiainen, Alam, Blackwell, Skarin, 2022), among others. An ensemble model consists of a finite set of independently trained alternative models that allow better performance than could be obtained from any of the individual models of the ensemble (Kunapuli 2023). The most commonly used ensemble model for classifying feeding activities are Adaptive Boosting (AdaBoost) (Wang, He, Zheng, Gao, Zhao, 2018; Carslake, Vázquez-Diosdado, Kaler, 2020), RF (Balasso, Marchesini, Ughelini, Serva, Andrighetto, 2021; Chang, Fogarty, Swain, García-Guerra, Trotter, 2022) and eXtreme Gradient Boosting (XGB) (Chen, Li, Guo, et al., 2022; Dutta, Natta, Mandal, Ghosh, 2022).

Classifiers based on DL methods include different types of Artificial Neural Networks (ANN) with hierarchical layers such as Multilayer Perceptron (MLP), CNN, Recurrent Neural Network (RNN) (Goodfellow, Bengio, Courville, 2016), and Long Short Term Memory (LSTM) (Hochreiter & Schmidhuber, 1997). Although DL methods are still less used than classic ML methods (Peng et al., 2019; Pavlovic et al., 2021; Hai, 2022; Petranović, 2022), its use as a classification model has increased recently because of its success in other applications. One distinctive feature of these models is their ability to process the raw data without feature engineering.

Supervised ML methods learn a function that maps features (inputs) to labels (output) based on example input-output pairs. The most widely used learning algorithms include k-NN (Dutta, Smith, Rawnsley, Bishop-Hurley, Hills, 2014; Bishop-Hurley et al., 2014; Sakai, Oishi, Miwa, Kumagai, Hirooka, 2019), Linear Discriminant Analysis (LDA) (Nielsen, 2013; Yoshitoshi et al., 2013), SVM (Vázquez-Diosdado et al., 2015), DT (Riaboff et al., 2019; Chebli, El Otmani, Cabaraux, Keli, Chentouf, 2022) and ANN (Chang, Fogarty, Swain, García-Guerra, Trotter, 2022). Unsupervised ML methods learn patterns from untagged data. The goal is to build a concise representation of the problem through machine output imitation and then generate imaginative content from the machine. The k-means classification has been successfully employed with accelerometers (Vázquez-Diosdado et al., 2019).

### 3.1.5. Validation methodology

Model validation is the process of evaluating a trained model, on a validation dataset, using a performance metric that indicates its generalisation capability. The validation data set provides an unbiased evaluation of a model fitted on the training data set while tuning the model's parameters.

The most popular technique for generating validation data sets is *k*-fold CV (Bishop-Hurley et al., 2014; Vázquez Diosdado et al., 2015; Barwick et al., 2018), being 5 (Riaboff et al., 2019; Hu et al., 2020) and 10 (Mansbridge et al., 2018; Hamilton et al., 2019) the most frequent values of *k*. Dataset segmentation into training and testing/validation sets was exploited by several authors using different ratios (Nielsen, 2013; Martiskainen et al., 2009; Alvarenga et al., 2016; Li, Cheng, and Cullen, 2021; Pavlovic et al., 2021). These previous two approaches are combined by creating an initial partition between training and testing, and then using a k-fold CV over the training partition to validate the model (Pavlovic et al., 2022; Li et al., 2022). Several authors explored model training and testing with sets of animals, implementing leave-one-animal-out (Rahman et al., 2018; Fogarty, Swain, Cronin, Moraes, Trotter, 2020; Arablouei et al., 2021) and leave-several-animal-out variant (Rahman et al., 2016) approaches.

The second methodological issue to consider is the metrics used to monitor and measure the performance of a model during training and validation. The most widely used are accuracy, precision, recall (sensitivity), specificity, and F1-score (Nielsen, 2013; Yoshitoshi et al., 2013; Guo, Welch, Dobos, Kwan, Wang, 2018; Mansbridge et al., 2018). Less frequently selected metrics are kappa (Martiskainen et al., 2009; González, Bishop-Hurley, Handcock, Crossman, 2015; Rolan et al., 2018; Barker et al., 2018), Matthew's correlation coefficient (Gonzales et al., 2015; Arablouei et al., 2021; Simanungkalit et al., 2021), the area under the curve (Cabezas et al., 2022), R2 (Rayas-Amor et al., 2017), misclassification rate (Tani, Yokota, Yayota, Ohtani, 2013), quality percentage, branching factor, and miss factor (Arcidiacono, Porto, Mancino, Cascone, 2017).

Finally, resampling techniques and metrics that prevent class bias are combined to address class imbalance problems. For example, Pavlovic et al. (2021) used a weighted F1-score, while Shen et al. (2021) analysed the results class by class.

### 3.2.   Acoustic sensors

Sounds produced during ruminants' feeding activities contain information about JM, feeding activities, and the type and amount of herbage intake and regurgitated. Thus, researchers develop specialised algorithms to extract this information from the sound: (i) individual event recognisers (JM recognition), (ii) continuous activity recognisers (rumination and grazing recognition), and (iii) parameter estimation algorithms (DMI, type of herbage). Since the pioneering work of Alkon and Cohen (1986), acoustic monitoring has become a practical methodology for studying animal feeding behaviour. Laca, Ungar, Seligman, Ramey, and Demment (1992) instrumented inward-facing microphones on the forehead of steers to register louder and distinguishable feeding sounds, proving to be a more effective technique for discriminating subtle differences in feeding activities than previous devices or methods. Since then, it has been increasingly adopted as a research tool for studying different aspects of ruminant feeding behaviour (Galli, Cangiano, Demment, Laca, 2006; Galli, Cangiano, Milone, Laca, 2011; Lorenzón, 2022).

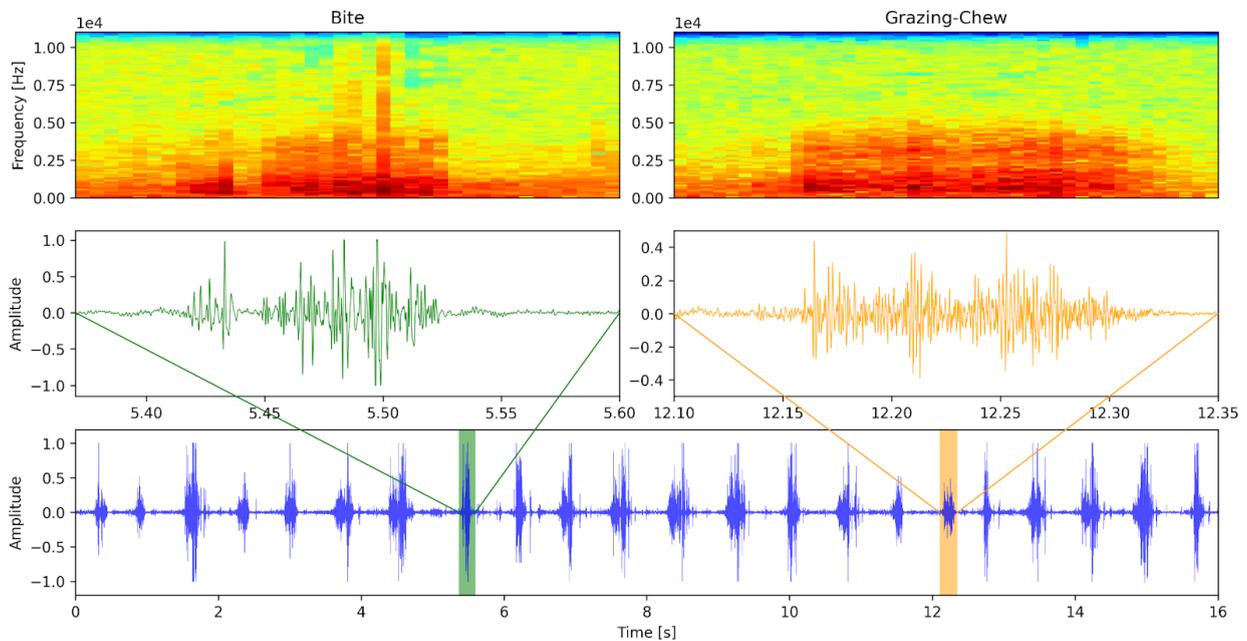

**Figure 10**: Sound signals recorded during grazing using a microphone on the cattle's forehead.

Figure 10 shows a typical sound record (and its time-frequency representation) of grazing cattle recorded using a microphone on the animal's forehead (Vanrell et al., 2020). It shows individual JMs (bite and grazing-chew). As has been demonstrated, there is a relationship between sound signals and the amount of dry matter ingested by the animal (Galli et al., 2018). However, sounds need to be processed to extract all this meaningful information.

### 3.2.1. Data acquisition and management

Acoustic monitoring faces challenges due to the scarcity of standardised and accessible datasets, with most studies relying on data collected by individual research teams and not shared with the broader research community. These datasets exhibit variations in experimental conditions such as the ruminant species, number of animals, observation period, grazing conditions, sensor types and locations, and pasture characteristics (type and height). Addressing the essential issue of data availability is crucial for further progress in this field.

While proprietary datasets remain prevalent in the literature, a notable exception is the audio dataset of ingestive JM made available by Vanrell et al. (2020). This dataset captures sounds produced by dairy cows during individual grazing sessions of tall and high fescue and alfalfa, recorded using microphones (Nady 151 VR, Nady Systems, Oakland, CA, USA) attached to the forehead of the cows and shielded with rubber foam, according to Milone, Galli, Cangiano, Rufiner, and Laca (2012). Comprising 52 raw audio signals in WAV format at 16-bits and 22.05 kHz, the dataset includes sequences of 3,038 JM events (bites, grazing chews, and chew-bites) and periods of silence contaminated with environmental noise.

Martinez-Rau et al. (2023) published a wider audio dataset comprising 708 hours of daily recordings acquired on five lactating multiparous Holstein cows for six non-consecutive days in both pasture and barn settings, registering 392 hours of grazing and rumination bouts. This dataset also includes two audio signals recorded during grazing and rumination sessions, respectively, containing more than 6,200 JM events (bites, grazing chews, rumination chews, and chew-bites). Audio signals were recorded in MP3 format using two electret microphones located in the forehead of the cows (Milone, Galli, Cangiano, Rufiner, Laca, 2012), connected to digital recorders (Sony Digital ICD-PX312, Sony, San Diego, CA, USA).

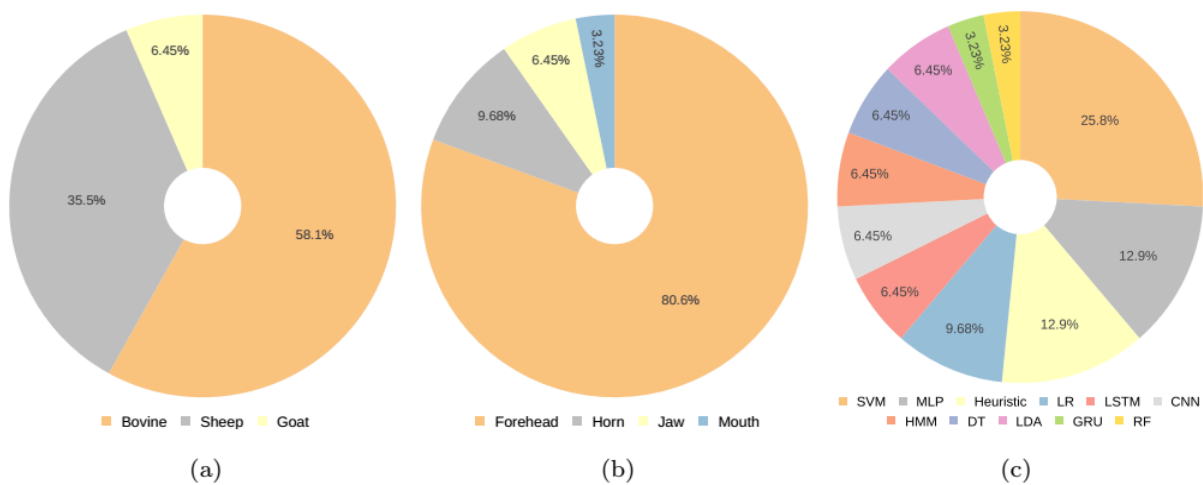

**Figure 11:** Ruminant species considered in the bibliography for acoustic monitoring (a). Location of sensors used for feeding behaviour monitoring based on sound (b). ML methods used for acoustic monitoring of ruminants (c).

Different ruminant species were employed in acoustic studies (Fig. 11.a). It shows that bovines are the most frequently used, almost two-thirds of all papers, followed by sheep with one-third of them. The contribution of goats studies to literature is minor, rising to only 6% of all works. This fact can be due to their economic significance and proportional population.

The works published in the literature analysed different grazing conditions, animal quantities, and observation periods. Some studies recorded data from animals confined in individual fenced plots (Duan et al., 2021; Sheng et al., 2020) or tie-stalls (Goldhawk, Schwartzkopf-Genswein,

Beauchemin, 2013). Others recorded data from animals bounded in loose indoor housing (Goldhawk, Schwartzkopf-Genswein, Beauchemin, 2013; Meen, Prior, Lam, 2016; Jung et al., 2021; Wang, Xuan, Wu, Liu, Fan, 2022; Li, Cheng, Cullen, 2021) or barns (Tani, Yokota, Yayota, Ohtani, 2013). Few studies recorded data from animals in free grazing conditions (Navon, Mizrach, Hetzroni, Ungar, 2013; Clapham, Fedders, Beeman, Neel, 2011; Wang, Wu, Cui, Xuan, Su, 2021; Chelotti et al., 2016; Vanrell et al., 2018; Chelotti et al., 2020), which is one of the most challenging scenarios. The number of animals employed in these experiments ranges from 3 to 225, while the observation period lasts from 5 hours to 25 days. These facts make it difficult to compare experimental results and comprehend the advantages and drawbacks of each algorithm.

Other technical conditions changing in the studies are the type of sensor and its location in the animal's body. In most cases, the devices are commercial wireless microphones (Ungar et al., 2006; Milone, Rufiner, Galli, Laca, Cangiano, 2009; Milone, Galli, Cangiano, Rufiner, Laca, 2012; Duan et al., 2021; Sheng et al., 2020; Wang, Wu, Cui, Xuan, Su, 2021; Wang, Xuan, Wu, Liu, Fan, 2022). In other cases, a commercial device (from SCR Engineers Ltd.) has been used for recording activities (Rodrigues et al., 2019; Goldhawk, Schwartzkopf-Genswein, Beauchemin, 2013). Few researchers have designed specific devices built upon open-hardware platforms (Deniz et al., 2017; Jung et al., 2021).

Most studies employ sensors attached to the animal's forehead (Ungar et al., 2006; Milone, Galli, Cangiano, Rufiner, Laca, 2012; Navon, Mizrach, Hetzroni, Ungar, 2013; Chelotti et al., 2016; Vanrell et al., 2018; Chelotti et al., 2020; Martinez-Rau et al., 2022). Tani, Yokota, Yayota, and Ohtani (2013) compared the performance in activity monitoring of cattle with sensors attached to the horn, nasal bridge, and forehead. Goats' and sheep's feeding behaviours have been monitored with piezoelectric microphones placed on the horns (Navon, Mizrach, Hetzroni, Ungar, 2013). Microphones are not unique sensors used to record sounds. A study has shown the effectiveness of a single-axis accelerometer in this task. It recorded the vibrations generated by animals during grazing and ruminating using a voice recorder (Tani, Yokota, Yayota, Ohtani, 2013).

The locations of the acoustic sensors were varied (Fig. 11.b). The most common mounting place is the forehead because it is easy to mount and provides direct information on JM, allowing recognition of feeding activities and estimation of forage intake. The other favoured places (jaw, mouth, and horn) are in the head, but the resulting signals have a lower SNR. They concentrate a small fraction (20%) of the studies, while the forehead concentrates the remainder (80%).

## 3.2.2. Preprocessing

Acoustic preprocessing methods are diverse and mostly influenced by those used in automatic speech recognition. Segmentation or windowing are typical strategies employed by acoustic monitoring algorithms. They allow the audio signal to be processed in real-time and at a low computational cost using fixed-length segments (Duan et al., 2021; Navon, Mizrach, Hetzroni, Ungar, 2013; Chelotti et al., 2016; Chelotti et al., 2020; Martinez-Rau et al., 2022). Most of these works use rectangular windows to define segments, while others use specific windowing such as sliding Hanning or Hamming windows (Sheng et al., 2020).

The SNR of the incoming audio signals is improved using different filters. In the literature, most of the algorithms employ linear time-invariant filters: high-pass (Clapham, Fedders, Beeman, Neel, 2011), fixed low-pass (Li, Cheng, Cullen, 2021; Tani, Yokota, Yayota, Ohtani, 2013; Navon, Mizrach, Hetzroni, Ungar, 2013; Chelotti et al., 2016), or notch filters (Galli, Cangiano, Milone, Laca, 2011). Specifically, notch filters remove band-limited noises and sounds introduced intentionally during the signal recording for synchronisation purposes. More robust algorithms are necessary to deal with time-varying and non-linear disturbances. In these cases, adaptive filters have been implemented with excellent results (Chelotti et al., 2018; Chelotti et al., 2020; Martinez-Rau et al., 2022).

### 3.2.3. Feature extraction

There is no clear agreement on the type of features (frequency-domain or time-domain) to use in the monitoring algorithms since both provide valuable information to achieve good classification results.

Mel-Frequency Cepstral Coefficients (MFCC) and their variants (log-scaled Mel-spectrogram representation) are the preferred frequency-domain features for the feature extraction stage (Deller, Hansen, Proakis, 2000). Its popularity lies in the fact that they have been a popular technique in automatic speech recognition, providing information to classify JMs (grazing chew, rumination chew, bite, and chew-bite) and estimate the amount of herbage processed (forage and dry matter intake (DMI)) by the animal. MFCC has been used to estimate forage intake in sheep (Sheng et al., 2020), and classify ingestive JM events in sheep (Millone et al., 2008; Galli et al., 2020; Duan et al., 2021) and dairy cows (Millone et al., 2012; Li, Cheng, Cullen, 2021). Tani, Yokota, Yayota, and Ohtani (2013) used time-frequency representations to classify grazing and rumination activities and count the total number of JMs.

Time-domain features are widely used because of their low computational cost, allowing real-time implementations in low-cost embedded systems (Deniz et al., 2017). Galli et al. (2020) highlighted their contribution to recognise JM events related to grazing behaviour. Time-domain features are computed from the conditioned sound signal segments, providing a *physical description of JM* through a set of intrinsic properties. They describe and quantify JM in terms of shape, duration, rate of change, maximum intensity, symmetry, and energy content (Fig. 12).

Clapham, Fedders, Beeman, and Neel (2011) used temporal and spectral features to identify bite events in free-grazing cattle. Navon, Mizrach, Hetzroni, and Ungar (2013) proposed a set of four temporal features to detect JMs, without distinguishing their type or class, in cattle, goats and sheep. Chelotti et al. (2016; 2018), and Martinez-Rau et al. (2022) used different sets of three to five temporal features to recognise JM events in dairy cows. Moreover, Chelotti et al. (2020) computed statistical features of recognised JM events to recognise grazing and rumination bouts in dairy cattle. Galli, Cangiano, Milone, and Laca (2011), Galli et al. (2018) and Lorenzón (2022) used temporal features of JM and LR models for DMI estimation in sheep and cattle. Based on the analysed literature, there is a tendency to use time-domain features.

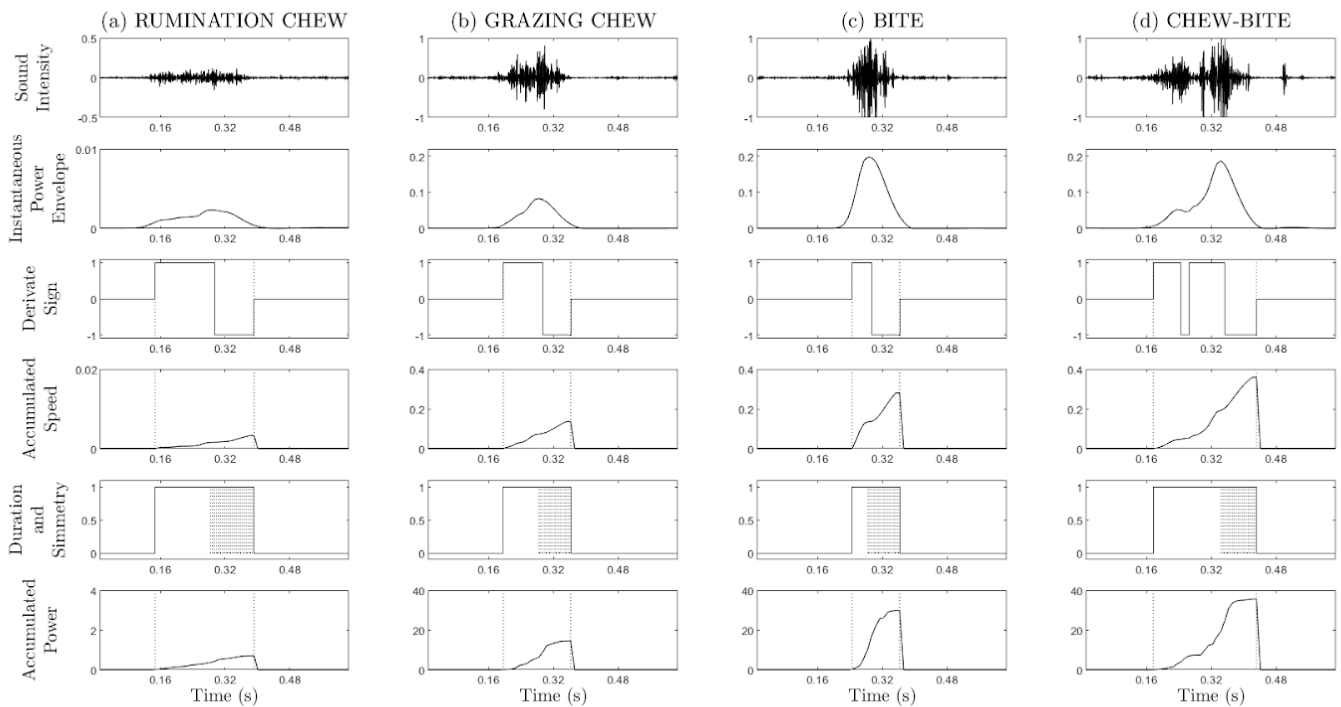

**Figure 12**: Acoustic signal during JM and corresponding distinctive features (adapted from Chelotti et al., 2016; 2018; and Martinez-Rau et al., 2022).

### 3.2.4. Classification

Different ML techniques have been reported to address problems related to ruminants' feeding behaviour, such as forage intake estimation, JM events classification, and feeding and rumination activities recognition. Clapham, Fedders, Beeman, and Neel (2011) classified bite events using rule-based analysis of the computed acoustic features. Galli, Cangiano, Milone, and Laca (2011) and Wang, Xuan, Wu, Liu, and Fan (2022) proposed LR models based on a set of explanatory variables computed from the chewing and biting sounds to estimate DMI in sheep. Similarly, Galli et al. (2018) used a LR model to estimate DMI in dairy cows. Sheng et al. (2020) proposed a classifier based on SVM to first identify chewing sound segments, and then estimate the forage intake using features extracted from detected JM in a least squares regression model combined with an elastic network.

Figure 11.c shows ML methods used for acoustic methods. Classic ML techniques comprise almost two-thirds (65.4%) of published works, followed by DL (19.2%) and heuristic (15%) models. The most widely used learning algorithms include SVM, MLP, DT, and RF (Bishop & Nasrabadi, 2006). They have been used to classify JM and feeding activities. DL methods include different types of ANN, including CNN and RNN. They can process the raw acoustic signal instead of working with the extracted features used by the heuristic and classic ML models. LR is the preferred statistical-based method used for estimating forage consumption and DMI (Galli, Cangiano, Milone, Laca, 2011; Galli et al., 2018; Wang, Xuan, Wu, Liu, Fan, 2022). On the other hand, heuristics methods use empirical rules and thresholds to discriminate JM and animal behaviours (Vanrell et al., 2018). Thresholds' values can be assigned manually derived from expert knowledge (Ungar et al., 2006; Clapham, Fedders, Beeman, Neel, 2011) or estimated from feature distribution (Chelotti et al., 2016).

The different types or classes of JM events can be detected and classified using a variety of approaches, ranging from heuristic rules to complex DL models. Tani, Yokota, Yayota, and Ohtani (2013) proposed an algorithm to identify cattle chewing activity based on the template-matching method applied to spectrogram segments. It distinguishes ingestive and ruminative JM without discriminating against individual JM. Milone, Rufiner, Galli, Laca, and Cangiano (2009) proposed four Hidden Markov Models (HMM) to classify JMs in sheep: the first one based on the acoustic level and linear prediction coefficients (LPC) as inputs; the second and third model coupled a sub-event level with an event level; and a compound model inspired by the language models widely used for speech recognition. Milone, Galli, Cangiano, Rufiner, and Laca (2012) built an acoustic model for classifying JM in dairy cows using HMM, filter-bank energies as features, and a long-term statistical model for capturing broad dependencies and constraints in possible JM event sequences. Galli et al. (2020) introduced an algorithm that uses a statistical classifier based on the LDA of LPC and a reduced set of spectral features. The common characteristic of the previous algorithms is that they were designed for *offline operation*, processing the full acoustic signal in a single step.

Alternatively, a series of acoustic algorithms have been developed for online operation, processing sample-by-sample or segments of the acoustic signal. Navon, Mizrach, Hetzroni, and Ungar (2013) discriminated JM from background noise. The algorithm used the level difference on the event sound envelope and noise segments to construct a maximum margin classifier. Chelotti et al. (2016) developed an algorithm that classifies individual JM (chew, bite, and chew-bite) in grazing cattle. It combined time-domain features computed from the sound envelope with heuristic rules. Its computational load allows real-time execution in low-cost embedded systems (Deniz et al., 2017). Chelotti et al. (2018) replaced the heuristic rules with classic ML techniques and enlarged the original set of features to improve the algorithm's performance. They also investigated the effect of different ML models (DT, RF, SVM, and MLP) on the system performance, without finding significant differences. Martinez-Rau et al. (2022) developed an algorithm that combines time-domain features with an MLP classifier. This work presented the first acoustic-based method for classifying four types of JM instead of three: three JM involved in grazing (grazing-chew, bite, and chew-bite) and one JM involved in rumination (rumination-chew).

Deep learning models have also proven to be highly effective. Li, Cheng, and Cullen (2021) proposed and compared different DL models for JM classification in cattle using sound. The models also analysed the effect of pasture heights on sounds. Their models combined 1D- and 2D-CNN with LSTM models. Wang, Wu, Cui, Xuan, and Su (2021) tackled the same problem for sheep using CNN and Gated Recurrent Units (GRU). Duan et al. (2021) proposed another algorithm based on LSTM networks for feeding event classification. The sound related to the events was isolated using a segmentation method based on short-term energy and average zero-crossing rate thresholds. A discrete wavelet transform-based MFCC feature, dimensionally reduced using principal component analysis, was used to train the neural network. The algorithm has successfully classified bite, ingestion-chew, bolus-regurgitation, rumination-chew, and unrelated-behaviour categories. Jung et al. (2021) presented a DL model for real-time classification of behavioural sounds of cattle. The sounds include feeding-related vocalisations like food-anticipating calls. The algorithm uses a 2D-CNN for identifying cattle vocals and removing background noises and a similar convolutional model to perform behaviour classification. Both models use MFCC as input.

While most of the authors focused on the recognition of JM events and the estimation of forage intake, acoustic methods have also been developed for classifying grazing and rumination activities. Vanrell et al. (2018) proposed an algorithm based on statistical information on sound signals to recognise feeding activities. It has two stages: segmentation and classification. The segmentation stage uses the regularity patterns of masticatory events to break down the sound record into segments. These regularity patterns are detected using the autocorrelation of the sound envelope. Then, the classification stage analyses the sound envelope energy to detect pauses and characterise their regularity. Chelotti et al. (2020) proposed an ML approach for grazing and rumination classification. It used a set of statistical features of recognised JM, analysed with an ML model, to recognise feeding activity bouts. This algorithm achieved a higher performance than that achieved by Vanrell et al. (2018), having a low computational load and being feasible for real-time implementation for online monitoring of foraging behaviour.

### 3.2.5. Validation methodology

Depending on the objective (JM recognition, activities classification or DMI estimation), each algorithm needs specific metrics and methods to be evaluated. Similar to motion-based sensors, the most popular validation technique for acoustic-based algorithms is $k$-fold CV (Galli et al., 2018; Galli et al., 2020; Wang, Xuan, Wu, Liu, Fan, 2022). The leave-one-signal-out approach is employed when multiple acoustic signals are available (Chelotti et al., 2018; Martinez-Rau et al., 2022). A simple separation into training and validation was also used in some works (Chelotti et al., 2016; Vanrell et al., 2018). Wang, Wu, Cui, Xuan, and Su (2021) and Li, Cheng, and Cullen (2021) separated the dataset into training, validation and test sets, while Chelotti et al. (2020) divided the dataset into two sets, one for training and validation using k-fold CV, and the other one for testing. At this point, it is important to emphasise that many works do not provide details of the hyperparameters tuned during the training and validation process.

A second methodological issue to analyse is the metrics used to monitor and measure the model performance during training and testing. For the recognition of JM events such as chew, bite, and chew-bite (Millone et al., 2009; Millone et al., 2011; Clapham, Fedders, Beeman, Neel, 2011; Tani, Yokota, Yayota, Ohtani, 2013; Navon, Mizrach, Hetzroni, Ungar, 2013; Chelotti et al., 2016; Galli et al., 2020) the authors used simple metrics such as accuracy, recognition rate, false positives, and false negatives to report their results. In recent years, many studies have used a set of standard metrics, such as specificity, recall, precision, and F1-score (Chelotti et al., 2018; Sheng et al., 2020; Duan et al., 2021; Li, Cheng, Cullen, 2021; Martinez-Rau et al., 2022). Among the advantages, this approach obtains more robust results regarding the data imbalance.

Measuring the performance of a feeding activity recogniser implies a particular challenge due to its continuous nature (Ward, 2011). Unlike discrete events, activity recognition requires the recognition of categories and the partial overlaps between the reference and the recognised sequences. In this sense, Vanrell et al. (2018) and Chelotti et al. (2020) addressed this problem using spider plots to provide a multi-dimensional analysis. Moreover, these diagrams presented both frame and block-based metrics, allowing us to analyse the activities recognition at different temporal scales. Studies addressing the DMI estimation evaluated the algorithm performance using standard metrics for regression such as R2 or MSE (Galli, Cangiano, Milone, Laca, 2011; Galli et al., 2018, Wang, Wu, Cui, Xuan, Su, 2021).

Most of the analysed works applied, resampling techniques and metrics that prevent class bias to tackle class imbalance problems.

### 3.3.    Image sensors

Although wearable sensors (Fig. 5) offer precise information, they have several limitations. They can be easily damaged, cause animal stress and discomfort (Kuan, Tsai, Hsu, Ding, Te Lin, 2019), and have limited autonomy (Farooq, Sohail, Abid, Rasheed, 2022). Furthermore, due to their specific location on the animal's body, wearable sensors often face compromises when tracking several behaviours simultaneously (Li, Jiang, Wu, Yin, Song, 2019).

The approaches based on computer vision are non-invasive, offer a high-speed response, and can avoid stress problems caused by wearable sensor monitoring. Cameras collect images since they are easy to deploy, providing a complete real-time understanding of the livestock farming scene. So, computer vision is an emergent development direction to improve animal behaviour recognition and analysis (Wu et al., 2021).

Image sensors have received increasing attention in the academic community, particularly in the last ten years. This interest arises from the availability of low-cost cameras and communication devices and the latest developments in image-processing methodologies (Chen, Dongjian, Yinxi, Huaibo, 2017; Porto, Arcidiacono, Anguzza, Cascone, 2015). Although image sensors have been used to estimate feed intake and classify animal activities and behaviours, none of the studies analysed focused on the recognition of JM events.

The body measurements of a ruminant are important characteristics to monitor, since they are closely related to its nutritional status and health. In this sense, methods and devices using 3D cameras have gained great popularity due to improvements in image quality and processing techniques in recent years (Du et al. 2022; Luo, Hu, Gao, Guo, Su, 2023). Several studies have used ML (particularly DL techniques) and similar computational approaches to assess the body condition score from 3D images, obtaining performance rates of approximately 75% or higher (Alvarez et al. 2018; Song, Bokkers, Van Mourik, Koerkamp, Van Der Tol, 2019; Liu, He, Norton, 2020; Zhang et al., 2023).

### 3.3.1.    Data acquisition and management

Like the other sensing techniques, the development of image-based solutions also faces the problem of the lack of accessible and standardised databases, hindering the evaluation and comparison of algorithms. Thus, each study uses its dataset, except for works presented by the same team of researchers.

A wide variety of experimental conditions have been considered in the literature, including the ruminant species (bovine (Ayadi et al., 2020), goat (Jiang, Rao, Zhang, Shen, 2020) and sheep (Deng et al., 2021)), the number of animals (ranging from 3 (Shiiya, Otsuka, Zin, Kobayashi, 2019) to 46 (McDonagh et al., 2021)), the position of cameras, and the observation period (amount of images (Yu et al., 2022) or period of time (Guo, Qiao, Sukkarieh, Chai, He, 2021)). Some studies recorded data from animals in fenced plots (Qiao, Guo, Yu, He, 2022; Guo, Qiao, Sukkarieh, Chai, He, 2021) or paddocks (Yin, Wu, Shang, Jiang, Song, 2020; Nguyen et al., 2021; Wu et al., 2021).

Other studies focus on free-stall barns (Yu et al., 2022; Kuan, Tsai, Hsu, Ding, Te Lin, 2019), indoor pens (Chen, Li, Guo, et al., 2022; Li, Jiang, Wu, Yin, Song, 2019), and other indoor scenarios (Achour, Belkadi, Filali, Laghrouche, Lahdir, 2020; Ayadi et al., 2020; Chen, Dongjian, Yinxi, Huaibo, 2017, 2018; Fu, Fang, Zhao, 2022).

Figure 13.a shows the proportion of ruminant species employed in image and video studies. It shows that bovines are the most frequently used, 90.5% of all papers, followed by sheep and goats with 4.7% each. This fact can be due to their economic significance and availability.

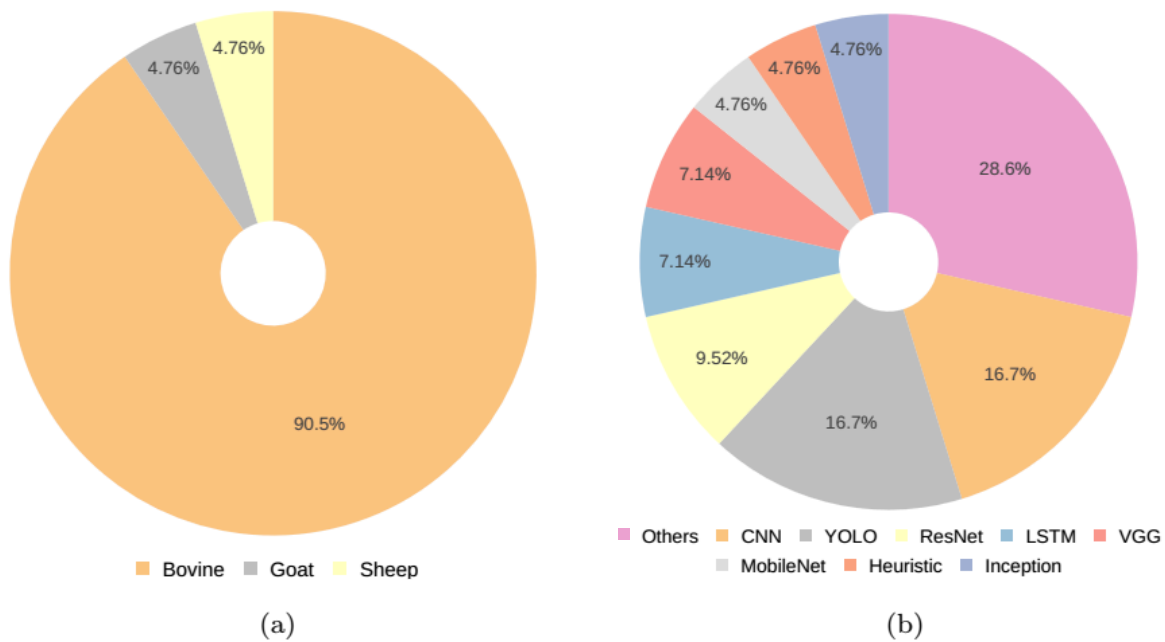

(a)                                    (b)

**Figure 13:** Ruminant species considered in the bibliography for image-based monitoring (a). Heuristic, classic ML methods, and DL models used for image and video monitoring (b).

The number of studies using ML with images and video is similar. The studies based on image sensors last from half an hour (Jiang, Rao, Zhang, Shen, 2020) up to six and a half hours (Wu et al., 2021; Nguyen et al., 2021). The studies based on video use different amounts of data, ranging from 247 images (Fu, Fang, Zhao, 2022) to 10288 (Yu et al., 2022). Most studies use 640x480 pixels images and videos, which are downsized before being used for model training (Achour et al., 2020; Ayadi et al., 2020). However, few studies use a higher video resolution: Guo, Qiao, Sukkarieh, Chai, and He (2021) used 704x576 pixels at 25 frames per second, while Li, Jiang, Wu, Yin, and Song (2019) used 1440×1080 pixels at 30 frames per second.

Training video-based algorithms requires more data than image-based ones, even at low frame rates. This fact stimulates data augmentation techniques to improve models' accuracy and robustness. Examples of these techniques are the random variations of the brightness in the Hue, Saturation, Value colour space, and rotations up to 25 degrees to make models invariant to the different postures (Kuan, Tsai, Hsu, Ding, Te Lin, 2019). In this sense, other operations for data augmentation include random flipping, random clipping, random rotation, and random scaling (Deng, 2021).

One aspect that most of these studies have in common is the fixed position of cameras, capturing the animals from a certain distance. Figure 14 shows different camera locations used in the bibliography. In most cases, there is one camera located in height: Shiiya, Otsuka, Zin, and Kobayashi (2019) used a directional camera (Fig. 14.a), and Wu et al. (2021) used a dome webcam (Fig. 14.b). Some studies used multiple cameras to prevent occlusion problems. Nguyen et al. (2021) used three cameras set on the top, the left, and the right of the area under study (Fig. 14.c). Yu et al. (2022) used two ZED2 binocular cameras (Stereolabs Inc., San Francisco, CA, USA), one placed on top of animals and another settled in front of them (Fig. 14.d). Qiao, Guo, Yu, and He (2022) used two frontal cameras in different locations for recording individualised calf and adult cows (Fig. 14.e). Finally, a multi-camera video-recording system of ten Vivotek FD7131 cameras (Vivotek Inc., New Taipei City, Taiwan) was proposed to obtain panoramic top-view images of the area under study (Porto, Arcidiacono, Anguzza, Cascone, 2015).

These studies used different types of cameras. However, the dome IP camera DS-2DM1-714 by Hikvision (Hangzhou Hikvision Digital Technology Co., Ltd., Hangzhou, Zhejiang, China) is the most frequently used (see Guo, Qiao, Sukkarieh, Chai, He, 2021; Chen, Dongjian, Yinxi, Huaibo, 2017; Chen, He, Song, 2018; Jiang, Rao, Zhang, Shen, 2020) because of its low cost, simple operation, installation, and maintenance.

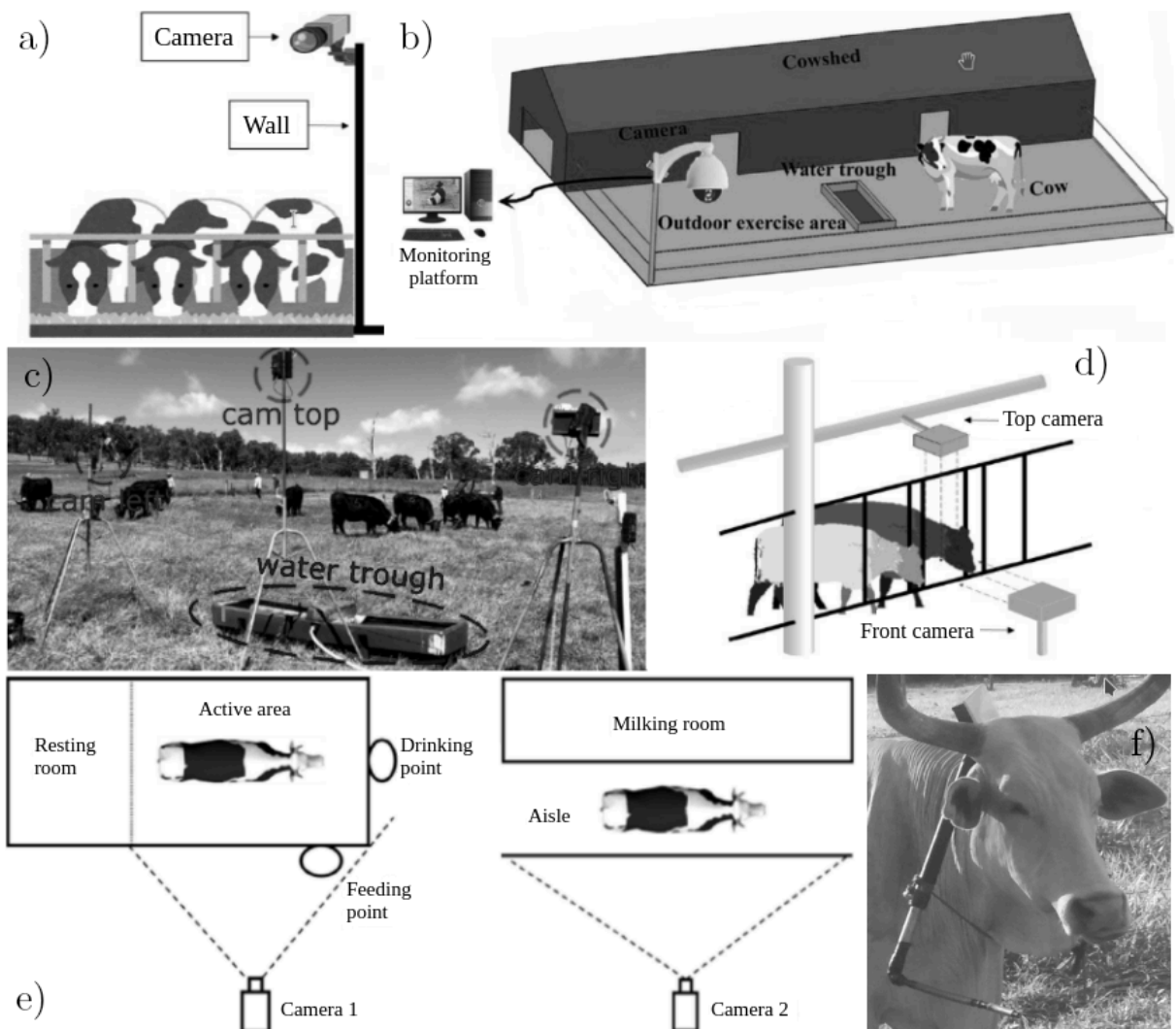

**Figure 14:** Camera locations used in the bibliography for capturing animal behaviour images and videos (adapted from (Shiiya et al., 2019; Wu et al., 2021; Nguyen et al., 2021; Yu et al., 2022; Qiao et al., 2022; de Oliveira et al., 2020)).

Studies that use cameras mounted on the animal's body are rare. De Oliveira et al. (2020) proposed a device attached to a cattle's neck to provide a close look at the mouth of the animal. It has a structural backbone with two portable cameras to capture frontal videos (during grazing) and lateral videos for observing the food bolus passing through the oesophagus (Fig. 14.f). These wearable cameras are often uncomfortable for the animal and may interfere with its natural behaviour.

### 3.3.2. Preprocessing and feature extraction

Images are usually captured with high-quality sensors under controlled lighting conditions, facts that reduce the impact of noise. Moreover, occlusions and illumination conditions are considered during image acquisition to improve the robustness of the models (Jiang, Rao, Zhang, Shen, 2020; Deng et al., 2021).

In traditional computer vision approaches, preprocessing steps such as image normalisation and filtering, and feature extraction are often necessary to extract meaningful information from images (Jingqiu, Zhihai, Ronghua, Huarui, 2017). ML-based approaches hardly perform preprocessing on images and videos since algorithms can usually capture the relevant information (Koohzadi & Charkari, 2017; Chen, Li, Bai, et al., 2021). Sometimes, it is required to improve algorithm performance and robustness. For example, the ML-based approach proposed by Porto, Arcidiacono, Anguzza, and Cascone (2015) calibrates, rotates, and resizes the images based on snapshots. Then, they are blended to obtain an output image to cover the area of interest. In a DL-based approach, preprocessing consisting of histogram equalisation was performed to improve the quality of the images by enhancing contrast (Kuan, Tsai, Hsu, Ding, Te Lin, 2019). Achour, Belkadi, Filali, Laghrouche, and Lahdir (2020) performed motion detection and background subtraction to compute a similarity index of consecutive images based on relevant images selected for model training.

Features extraction stages based on learning models provide a simpler processing pipeline and better model performances. Most of the developments based on images are built upon DL neural networks, using convolutional layers to perform feature extraction at different levels (Ayadi et al., 2020; Bezen, Edan, Halachmi, 2020; Qiao, Guo, Yu, He, 2022; Guo, Qiao, Sukkarieh, Chai, He, 2021; Yu et al., 2022; Kuan, Tsai, Hsu, Ding, Te Lin, 2019; Jiang, Rao, Zhang, Shen, 2020). DL models automatically extract relevant features from the raw or preprocessed image data without the need for manual feature engineering (Nguyen et al., 2021; Chen, Li, Guo, et al., 2022; Fu, Fang, Zhao, 2022; Deng et al., 2021; McDonagh et al., 2021; Shang, Wu, Wang, Gao, 2022).

Achour et al. (2020) proposed a feature extraction stage based on four convolutional and pooling layers. Yin, Wu, Shang, Jiang, and Song (2020) used an efficient DL model based on EfficientNet to extract spatial features from videos of cow behaviour. EfficientNet is a CNN model with high parameter efficiency and speed (Koonce 2021). In this model, the features of the first layers provide information about textures and edges, being susceptible to interference because of the complex background of cattle farms (Jeong, Park, Henao, Kheterpal, 2023). Thus, the size of these feature maps becomes boundless, increasing the model complexity and computational time. Then, the authors proposed a multilevel fusion of features using a bidirectional feature pyramid network (Cao, Zhang, Zhong, 2021) to overcome this problem.

### 3.3.3. Classification

Figure 13.b shows the frequency of heuristic, classic ML methods and DL models used for image and video analysis. Heuristic methods represent only a small fraction of 4.8% of the studies (Li, Jiang, Wu, Yin, Song, 2019; Shiiya, Otsuka, Zin, Kobayashi, 2019). In contrast to wearable sensors, most algorithms based on images and videos employ DL for their implementation. CNN-based models (CNN, You-Only-Look-Once (YOLO), ResNet, VGG, MobileNet, and Inception) represent almost two-thirds of the studies (59.6%). CNNs can achieve outstanding performances on a wide range of classification problems, being the most successful DL technique for image and video classification tasks. CNNs can automatically learn hierarchical representations by concatenating convolutional, pooling and flatten layers followed by ANNs. Therefore, they can effectively uncover spatial relationships and local patterns within images, making them particularly well-suited for object recognition, scene classification, and image/video classification.

To the best of the authors' knowledge, only seven publications were published using computer vision and classic ML methods, representing 28.6% of the image sensor-based studies (included in "Others" in Fig.13.b). Porto, Arcidiacono, Anguzza, and Cascone (2015) developed an algorithm for cow feeding and standing classification based on the Viola-Jones object detection framework. It uses Haar-like features and an ensemble classification approach called AdaBoost (Ying, Qi-Guang, Jia-Chen, Lin, 2013; Wang, 2014). Chen, Dongjian, Yinxi, and Huaibo (2017) introduced an algorithm based on the Mean Shift Tracking (MST) framework to detect cow rumination behaviour. The MST is a non-parametric estimation method for clustering, tracking, segmentation, and image smoothing (Dong and Catbas, 2021). Lately, Chen, He, and Song (2018) also introduced a target tracking framework, known as Spatio-Temporal Context learning, to solve the same problem. Li, Jiang, Wu, Yin, and Song (2019) presented an approach for tracking multiple ruminant mouth areas based on Horn-Schunck and Inter-Frame Difference algorithms. The Horn-Schunck algorithm estimates the motion (Dong and Catbas, 2021), while the Inter-Frame Difference algorithm discriminates between foreground and background by analysing consecutive frames. The authors used the Horn-Schunck algorithm to automatically detect cows' mouth areas, while the Inter-Frame Difference algorithm to track each cow's mouth area. Shiiya, Otsuka, Zin, and Kobayashi (2019) introduced a computer vision approach for cow feeding behaviour detection. It uses colour distance images to extract the cow region, computing the difference between frames, and then the feeding behaviour is determined using the extraction ratio and bounding box. Finally, Fuentes et al. (2022) proposed a regression algorithm based on MLP to estimate feed intake and rumination time, among other welfare targets, from video data.

De Oliveira et al. (2020) evaluated and compared different classic ML approaches (including SVM, RF, k-NN, and Adaboost) to analyse cows' mouth positions (mouth opened, closed, or intermediate) during rumination. The work also includes a performance comparison of several CNN-based models. It includes VGG16, VGG19, ResNet-50, InceptionV3, and Xception models. VGG16 and VGG19 are CNN models consisting of 16 and 19 layers of convolution, fully connected, MaxPool, and SoftMax operations. ResNet-50 is a residual neural network with 50 layers (Jeong, Park, Henao, Kheterpal, 2023). A residual network learns residual functions referenced to the layer inputs instead of unreferenced functions. These networks include skip connections (which perform identity mappings) merged with the outputs layer. InceptionV3 is a convolutional architecture from the Inception family based on depthwise separable convolution layers (Jeong, Park, Henao, Kheterpal, 2023). It uses label smoothing and an auxiliary classifier to propagate label information through the model. Similarly, Ayadi et al. (2020) tuned a pretrained VGG16 model using transfer learning (Weiss, Khoshgoftaar, Wang, 2016) to recognise rumination activity. They also compared the DL architecture's performance versus other CNN-based models (DenseNet, Inception, and ResNet). In another work, McDonagh et al. (2021) analysed video frame-by-frame with a ResNet-50 to classify cow activities like eating and drinking.

Other authors concatenated multiple single-task CNNs. Achour, Belkadi, Filali, Laghrouche, and Lahdir (2020) introduced an architecture based on four CNNs for monitoring the feeding behaviour of dairy cows. The first CNN detects the presence of a cow in the feeder zone. The second one determines the activity performed by the cow in the feeder. The third CNN checks the food availability and recognises the food category. The last CNN was coupled to an SVM to identify individual cows. Bezen, Edan, and Halachmi (2020) introduced an architecture based on two CNNs to estimate the intake of dairy cows. The first CNN identifies individuals based on the digits on their collars, while the second one estimates the feed intake.

The YOLO family consists of different YOLO models identified by version numbers. They have been used to develop classifiers in several studies on livestock monitoring. The YOLO model is a popular object detection framework known for its real-time performance and accuracy (Jiang, Rao, Zhang, Shen, 2020). The YOLO model implements a single-shot detection approach: one pass of the input data through the network to detect an object. YOLO's architecture consists of a CNN connected to a set of detection layers, which incorporates feature fusion at multiple scales to handle the different sizes of objects and capture the context and the details at different scales. Then, the feature maps passed through a series of detection layers responsible for predicting bounding boxes, object class probabilities, and confidence scores. YOLO architecture can perform multi-class object detection: predicts the probabilities corresponding to each object class for each bounding box.

Kuan, Tsai, Hsu, Ding, and Te Lin (2019) introduced an architecture based on two CNNs to estimate the intake of dairy cows. The first CNN, a Tiny-YOLOv2 (a YOLOv2 with fewer layers), detects the cow face, and the second one, a MobileNetV1, recognises the cow face. Jiang, Rao, Zhang, and Shen (2020) compared the performances of YOLOv3, YOLOv4, and faster Region-based CNN (R-CNN) InceptionV2 for goat activities classification. Results showed that YOLOv4 provides better real-time performance than the other models in speed detection and classification accuracy.

Yu et al. (2022) proposed a DL model to automatically identify feeding, chewing, and grass-bending behaviours in multiple cows. This architecture aims to track and quantify the feeding process and head movement trajectory in real-time. It is based on a YOLOv4 model with the addition of transformer enhancement modules (Chen, Li, Bai, et al., 2021). The reported results show improvements in feature extraction and monitoring accuracy. Deng et al. (2021) proposed a model based on YOLO to identify eating and postures in sheep. It uses a YOLOv3 model to extract the features and a pyramid feature fusion with a multi-scale prediction module for classification. Similarly, Shang, Wu, Wang, and Gao (2022) combined transformer modules with a MobileNetV3 model to obtain an architecture that improved the classification performance of standing, feeding, and lying activities. Furthermore, other studies have proven the advantages of YOLOv5 for the image classification of activities like drinking, feeding, standing, and lying in cows (Fu, Fang, Zhao, 2022) and sheep (Chen, Li, Guo, et al., 2022).

RNNs are often combined with CNNs to capture and exploit temporal information when analysing video. A bidirectional RNN integrates a forward and a backward RNN, capturing the hidden information from the past and future (Schuster & Paliwal, 1997). Yin, Wu, Shang, Jiang, and Song (2020) integrated an EfficientNet model with a bidirectional LSTM model, including an attention mechanism to classify cows' lying, standing, walking, drinking, and feeding activities. The EfficientNet extracts the features from each video frame, while the bidirectional LSTM is used to classify activities from the extracted features. Similarly, Wu et al. (2021) introduced a framework where the VGG16 model was used as the backbone to extract video feature sequences and a bidirectional LSTM for classification. This architecture provided better results than well-known models (VGG19, ResNet18, ResNet101, MobileNetV2, and DenseNet201) in activity classifications such as drinking, rumination, walking, standing, and lying. Following these ideas, Guo, Qiao, Sukkarieh, Chai, and He (2021) used an InceptionV3 to extract features from each video frame and a bidirectional GRU (BiGRU) (Yu, Si, Hu, Zhang, 2019) to extract spatial-temporal-features, incorporating an attention mechanism to keep the focus on key spatial-temporal-features. The classification results obtained in exploring, feeding, grooming, standing, and walking activities, show improvements compared to similar architectures without attention mechanisms.

Qiao, Guo, Yu, and He (2022) proposed an architecture that combined a 3D-CNN with a convolutional-LSTM module (Yu, Si, Hu, Zhang, 2019) to classify feeding activities. While standard LSTM models are unsuitable for modelling spatial data sequences (they only process one-dimensional data), the proposed architecture extends the convolution along the temporal direction to learn discriminative visual features and their temporal relations from the frames. Nguyen et al. (2021) used a cascade of R-CNNs (Cai & Vasconcelos, 2018) to detect cows, and a Temporal Segment Network (TSN) was used to classify activities. The TSN is a CNN that aims to model long-range temporal structures using a particular segment-based sampling and aggregation module (Koohzadi & Charkari, 2017).

### 3.3.4. Validation methodology

CV or multi-fold validation techniques are rare among works using images or videos (Oliveira,

Pereira, Bresolin, Ferreira, Dorea, 2021; Shang, Wu, Wang, Gao, 2022). The most common procedure to validate image- and video-based models uses a single data partition: training and validation datasets. The most common setup uses 80% of the data for training and the remaining 20% for validation. Some researchers separated images or video frames into different sets (Ayadi et al., 2020; Bezen, Edan, Halachmi, 2020), while others split completed video clips (McDonagh et al., 2021; Guo, Qiao, Sukkarieh, Chai, He, 2021; Qiao, Guo, Yu, He, 2022). Wu et al. (2021) slightly modified these percentages, keeping 30% of videos for validation and the remaining for training. These changes in the sizes of the training and validation datasets were extended to works using images (Porto, Arcidiacono, Anguzza, Cascone, 2015; Kuan, Tsai, Hsu, Ding, Te Lin, 2019; Achour et al., 2020; Deng et al., 2021; Fu, Fang, Zhao, 2022; Yu et al., 2022).

Using leave-one-out validation, Shiiya, Otsuka, Zin, and Kobayashi (2019) used five videos for evaluation and one for training. The idea behind this approach is to maximise the generalisation capabilities of the models. Some authors used an additional third test dataset to evaluate model performance, obtaining an indicator of generalisation capability and checking for possible overfitting. (Yin, Wu, Shang, Jiang, Song, 2020; Nguyen et al., 2021; Fuentes et al., 2022). Shang, Wu, Wang, and Gao (2022) initially employed a dataset for cow face detection and cow action classification. Subsequently, they utilised a secondary dataset to assess the model's generalisation capability across other livestock species such as pigs, sheep, and goats. Chen, Li, Guo, et al. (2022) divided the dataset using leave-one-animal-out rather than splitting fixed images.

Although most of the papers clearly describe all the elements for model training and validation (datasets and methodologies), there are few papers where this information is not detailed (Chen, Dongjian, Yinxi, Huaibo, 2017; Chen, He, Song,, 2018; Li, Jiang, Wu, Yin, Song, 2019; Jiang, Rao, Zhang, Shen, 2020).

There is no standardised methodology and tools for model evaluation and comparison of monitoring methodologies based on image sensors. The most basic and widespread metric for behaviour classification or animal recognition is accuracy (Nguyen et al., 2021; McDonagh et al., 2021; Shang, Wu, Wang, Gao, 2022). However, accuracy alone has limitations and can be misleading when the datasets are imbalanced. Besides, it treats all misclassifications equally, disregarding the potential consequences of the different types of errors. Due to these problems, studies incorporate other metrics besides accuracy for a more appropriate evaluation. Metrics like precision, recall, and F1-score are usually combined to achieve an accurate evaluation (Oliveira, Pereira, Bresolin, Ferreira, Dorea, 2021; Yin, Wu, Shang, Jiang, Song, 2020; Ayadi et al., 2020; Guo, Qiao, Sukkarieh, Chai, He, 2021; Fu, Fang, Zhao, 2022; Qiao, Guo, Yu, He, 2022; Cheng et al., 2022; Yu et al., 2022). Other commonly used metrics for behaviour event recognition are sensitivity (Porto, Arcidiacono, Anguzza, Cascone, 2015) and specificity (Wu et al., 2021). Confusion matrix is another powerful tool for performance analysis often used (Kuan, Tsai, Hsu, Ding, Te Lin, 2019; Achour, Belkadi, Filali, Laghrouche, Lahdir, 2020; Guo, Qiao, Sukkarieh, Chai, He, 2021; Qiao, Guo, Yu, He, 2022). It provides a detailed breakdown of the model's predictions for each class, allowing the identification of specific types of errors.

Feed intake estimation is another important task in this domain. The metrics considered for this problem are mean absolute error (Bezen, Edan, Halachmi, 2020), mean square error (Bezen, Edan, Halachmi, 2020; Fuentes et al., 2022), and correlation coefficient (Kuan, Tsai, Hsu, Ding, Te Lin, 2019; Fuentes et al., 2022).

Finally, a subproblem related to behaviour recognition and feed intake estimation is object detection. In this case, the objective is to detect the animal to be segmented and isolated from the background such that it is tracked in a video sequence to determine its activity. The metric used to evaluate the models developed for this task is the intersection over union measure (Kuan, Tsai, Hsu, Ding, Te Lin, 2019; Deng et al., 2021).

### 3.4. Other sensors

When cows feed, they move their jaws up and down, causing vibrations in the temporal bone. Movements can be sensed by measuring either the strain (pressure) changes on a rubber band (a tube filled with oil) mounted on the cow's nose (Fig. 15.a) or the vibrations in the temporal bone (Chen, Cheng, Wang, Han, 2020). Thus, noseband sensors directly sense JM (Fig. 15.b) (Dado & Allen, 1993; Rutter, Champion, Penning, 1997; Rutter, 2000; Kröger et al., 2016), providing relevant information for JM classification.

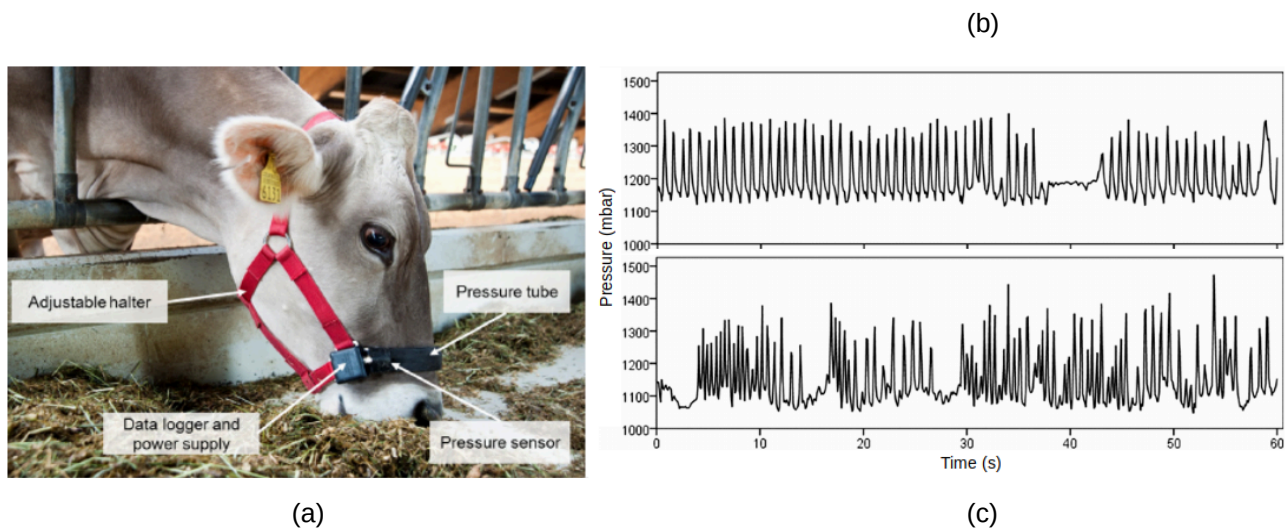

(b)

(a)                                        (c)

**Figure 15:** Technical components of the noseband sensor and raw signals recorded during b) rumination and c) grazing (adapted from Zehner, Umstätter, Niederhauser, Schick, 2017).

Noseband sensors require generating data for calibration and validation. This task is laborious, and the device's storage capacity and power supply limit the recording time (Nydegger et al., 2010). This type of sensor has been used for monitoring and assessing feeding activities (Werner et al., 2018; Li, Cheng, Cullen, 2021; Raynor, Derner, Soder, Augustine, 2021), health problems (Antanaitis et al., 2022), drinking activities during transition periods and lactation (Brandstetter, Neubauer, Humer, Kröger, Zebeli, 2019), peripartum period (Braun, Tschoner, Hässig, 2014) and calving (Fadul et al., 2022), among others.

### 3.4.1.  Data acquisition and management

One factor that hinders the development of pressure sensors is the difficulty of manufacturing. Nydegger et al. (2010) developed the first compact-built noseband pressure sensor system. This work establishes the basis for developing the commercial RumiWatch sensor system (Itin+Hoch GmbH, Liestal, Switzerland) designed for research purposes. The RumiWatch sensor includes a noseband pressure sensor (Kröger et al., 2016; Ruuska, Kajava, Mughal, Zehner, Mononen, 2016; Zehner, Umstätter, Niederhauser, Schick, 2017; Guccione et al., 2019; Li, Cheng, Cullen, 2021), optionally accompanied by an accelerometer located in the leg (pedometer) for measuring body motions and postures (Zehner et al., 2012; Werner et al., 2018; Poulopoulou et al., 2019). Most of the pressure-based studies employed the RumiWatch system for data acquisition. However, other authors developed their own pressure sensors. This system allowed individual JM recording but required animal-specific learning data. Chen, Cheng, Wang, and Han (2020) developed an activity sensor system based on an ultra-low power bubble activity sensor in the temporal fossa. Similarly, Chen, Li, Guo, et al. (2022) developed their noseband pressure sensor.

Another difficulty is the lack of standardised and accessible datasets. Most studies used datasets compiled by the research team, which are not generally available to the research community. Most of the analysed articles deal with the validation of the RumiWatch system in different grazing conditions (Werner et al., 2018), varying the number, species and age of animals (Eslamizad et al., 2018; Guccione et al., 2019), and the experimental periods. Some studies recorded data from animals confined in tie stalls (Braun, Trösch, Nydegger, Hässig, 2013). Others recorded data from animals bound in loose indoor housing (Ruuska, Kajava, Mughal, Zehner, Mononen, 2016; Kröger et al., 2016). Most studies recorded data in free-grazing conditions (Zehner, Umstätter, Niederhauser, Schick, 2017; Werner et al., 2018; Li, Cheng, Cullen, 2021). The number of animals employed in these experiments ranges from 3 (Chen, Li, Guo, et al., 2022) to 60 (Zehner,

Umstätter, Niederhauser, Schick, 2017), the experimental period goes from half a day (Guccione et al.,2019) to 30 days, and recording periods range from 100 minutes (Guccione et al.,2019) to 403 hours (Ruuska, Kajava, Mughal, Zehner, Mononen, 2016). These facts make it difficult to compare experimental results and comprehend the advantages and drawbacks of each algorithm (Pereira, Sharpe, Heins, 2021).

### 3.4.2. Preprocessing and feature extraction

The range of raw pressure data varies significantly between individual animals, and such scale difference affects the data modelling (Singh & Singh, 2020). Data preprocessing techniques eliminate this scale difference and normalise the scale. Data collected from cattle have different initial pressures (generated after wearing the noseband) because of the differences in cattle heads. This initial pressure value is a relatively stable constant during device operation. There are two ways to eliminate it: one is to extract local changes in the data, and the other is signal filtering. Some authors used first-order difference and local slope to extract local variation of data (Chen, Cheng, Wang, Han, 2020; Chen, Li, Guo, et al., 2022). A high-pass filter was used to remove unstable initial variables (Chen, Li, Guo, et al., 2022).

The segmentation stage divides the conditioned signals into fixed-length segments (*windows*) of arbitrary fixed values of 1 minute with 10 seconds overlapped (Braun, Trösch, Nydegger, Hässig, 2013; Zhener et al., 2017; Benaissa, Tuyttens, Plets, Cattrysse, et al. 2019; Chen, Cheng, Wang, Han, 2020; Chen, Li, Guo, et al., 2022). Some authors used larger windows to consolidate the partial estimates (5 minutes (Braun, Trösch, Nydegger, Hässig, 2013), 10 minutes (Zehner, Umstätter, Niederhauser, Schick, 2017; Norbu et al., 2021), and 60 minutes (Zhener et al., 2017; Steinmetz, von Soosten, Hummel, Meyer, Dänicke, 2020).

Time-domain features are the most frequently used with noseband sensors because of their low computational cost (Nydegger et al., 2010; Chen, Li, Guo, et al., 2022). They are computed from the conditioned pressure signal segments using statistics and signal processing. They describe the JM through a set of physical properties that describe them (rate of change, maximum amplitude, event period, inter-event period, and local slope), as well as a set of statistics (average, variance, and standard deviation).

Statistical characteristics of the frequency representation (mean, standard deviation, and correlation) are also computed as features (Chen, Cheng, Wang, Han, 2020). Some authors use spectral data like the fundamental frequency (Chen, Cheng, Wang, Han, 2020) and specific bands (Chen, Li, Guo, et al., 2022).

### 3.4.3. Classification

Heuristics methods are the most popular classification methods used by pressure-based noseband sensors. Data collected with the RumiWatch sensor are processed with proprietary software to discriminate JM events (Werner et al., 2018, Li, Cheng, Cullen, 2021) and animal behaviours (Nydegger et al., 2010; Braun, Trösch, Nydegger, Hässig, 2013). The software uses simple empirical rules derived from expert knowledge to evaluate feature values (Zehner, Umstätter, Niederhauser, Schick, 2017) (Fig. 16). Benaissa, Tuyttens, Plets, Cattrysse, et al. (2019) proposed a method that utilises DT and SVM algorithms to recognise feeding and rumination activities. They employed data collected from the RumiWatch and a neck-mounted accelerometer, achieving similar performance with each sensor.

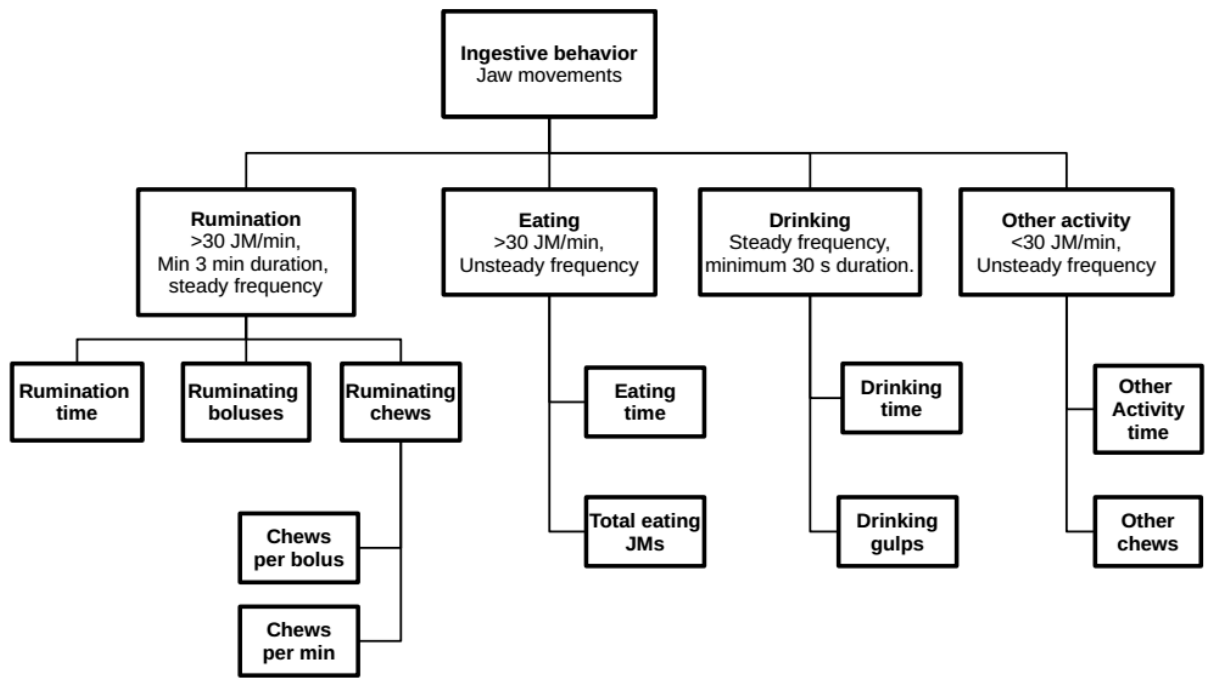

**Figure 16:** Classification tree of ingestive behaviours applied by the RumiWatch algorithm (Zehner, Umstätter, Niederhauser, Schick, 2017).

Regarding self-developed pressure sensors, Chen, Cheng, Wang, and Han (2020) compared the performance of ANN, RNN and CNN to identify the feeding behaviour of dairy cows. However, this method requires manual sensor calibration. Chen, Li, Guo, et al. (2022) proposed a classic ML approach using the XGB algorithm to eliminate the influence of the initial pressure of noseband sensors on rumination and eating behaviour identification. The method mainly used the local slope to obtain the local data variation and combined it with the Fast Fourier Transform to extract the frequency-domain features.

### 3.4.4. Validation methodology

As previously mentioned, some authors validated the performance of a commercial noseband pressure in particular animal or grazing conditions. Ground-truth references are generated by visual observation (Braun, Trösch, Nydegger, Hässig, 2013), sometimes assisted by cameras (Zehner et al., 2012). Zehner et al. (2012) measured the performance for counting the number of JM produced during eating and rumination using the mean absolute percentage error and the standard error of the mean. Braun, Trösch, Nydegger, and Hässig (2013) used a statistical test to discover statistically significant differences in the number and duration of individual rumination, eating and resting phases, the total daily length of these phases, the number of regurgitated cuds per day, the number of chewing cycles per cud and the total daily number of chewing cycles during eating and rumination. Kröger et al. (2016) studied variation in diets and discovered significant differences in several chewing variables using the analysis of variance (ANOVA) test and the concordance correlation coefficient (CCC) metrics. Ruuska, Kajava, Mughal, Zehner, and Mononen (2016) proposed a random coefficient regression model discovering systematic errors in eating and drinking behaviours in dairy cows. Similarly, Eslamizad et al. (2018) also used a random coefficient regression model in calves. Zehner, Umstätter, Niederhauser, Schick (2017) and Poulopoulou et al. (2019) used the Spearman correlation coefficients to measure the device performance for different behaviours of stable-fed cows and grazing beef cattle, respectively. Werner et al. (2018) evaluated the grazing, rumination, walking, standing, and lying duration per hour using the CCC in a pasture-based system. The Cohen's Kappa coefficient metric has been used to analyse the number of bites and rumination chew events (Werner et al., 2018), among other characteristics of the feeding behaviours (Guccione et al., 2019). Li, Cheng, and Cullen

(2021) used the CCC metric to evaluate the confidence level for quantifying and differentiating prehension bites, eating chews, and rumination chews events. Finally, Steinmetz, von Soosten, Hummel, Meyer, and Dänicke (2020) classified several behaviours at 1-minute and 1-hour scales using typical metrics (sensitivity, specificity, precision, accuracy, and Matthews Correlation Coefficient).

A train/validation split of the dataset was performed in studies developing their classification algorithms. Nydegger et al. (2010) used the train data to adjust the threshold values and heuristic rules to count JMs associated with different behaviours. The authors used the validation data to assess the performance using the percentage error metric. Chen, Cheng, Wang, and Han (2020) used 75% of the dataset to train the CNN model and the remaining 25% to evaluate the accuracy in recognising feeding behaviours.

Other authors used a CV strategy for training and evaluating the algorithms. While Chen, Li, Guo, et al. (2022) split the dataset into 5 folds, Benaissa, Tuyttens, Plets, Cattrysse, et al. (2019) used leave-one-animal-out. They measured the performance using precision, sensitivity, specificity, and F1-score.

## 4. Commercial devices

Commercial devices for cattle monitoring have been available on the market since the last decade of the previous century. These devices can distinguish behaviours associated with feeding, drinking, postures, locomotion, physical condition, and health (Stygar et al., 2021). Typically, commercial sensors have two parts: a data-logger acquisition system and a data analysis software tool. The software runs proprietary algorithms to report the information output. The lack of technical information about the algorithms and the validation procedures has motivated the development of alternative software. However, processing the raw data recorded by a commercial data logger is no longer feasible and depends on the sensor model.

More than a hundred retailed systems for animal-based welfare assessment are available in the market. Only 14% of the systems have been validated by groups different from the one that developed. Systems based on accelerometers are the most certified (30% of tools available on the market), while systems based on cameras and boluses are less validated (10% and 7% respectively). Validated attributes focused on animal activity, feeding and drinking behaviours, physical condition, and animal health. The majority of these systems have been verified on adult cows. Non-active behaviour (lying and standing) and rumination were the most often validated. The precision and accuracy of feeding and drinking assessment varied depending on measured traits and the used sensor. Table 2 summarises the most widespread technologies for monitoring feeding-related activities.

Table 2: Technologies for monitoring feeding-related activities (adapted from Stygar et al. 2021).

| Technology (provider) | Reference | Measured traits | Used sensor and attached position |
|---|---|---|---|
| Growsafe (GrowSafe Systems Ltd., Airdrie, AB, Canada) | DeVries, Von Keyserlingk, Weary, and Beauchemin (2003) | Presence at the feeder | RFID (neck collar), load cell |
| Insentec (Insentec, Marknesse, the Netherlands) | Chapinal, Veira, Weary, and von Keyserlingk (2007) | Presence at the feeder; Feed intake | RFID (ear), load cell |
| Hi-Tag (SCR Engineers Ltd., | Schirman et al. (2009) | Rumination time | Microphone, collar |
| Netanya, Israel) | | | |
| Ice Tag (IceRobotics Ltd., Edinburgh, Scotland) | Mattachini, Riva, Bisaglia, Pompe, and Provolo (2013) | Lying and standing behaviours | Accelerometer, leg |
| CowManager SensOor (Agis, Harmelen, Netherlands) | Bikker et al. (2014) | Lying and standing time; Rumination time | Accelerometer, ear |
| Intergado (Intergado Ltd., Contagem, Minas Gerais, Brazil) | Chizzotti et al. (2015); Oliveira Jr et al. (2018) | Presence at the feeder; Feed intake | RFID (ear), load cell |
| Smartbow (Smartbow GmbH, Jutogasse, Austria) | Borchers, Chang, Tsai, Wadsworth, and Bewley (2016) | Rumination time | Accelerometer, ear |
| RumiWatch (Itin+ Hoch GmbH, Liestal, Switzerland) | Zehner, Umstätter, Niederhauser, and Schick (2017); Werner et al. (2018) | Lying and standing time; Feeding time; Grazing and rumination time | Accelerometer and pressure sensor, halter and leg |
| MooMonitor+ (Dairymaster, Tralee, Ireland) | Werner et al. (2019) | Grazing and rumination times | Accelerometer, collar |

In the literature reviewed, studies differ in the commercial sensor employed as a data logger. The choice depends on the sensing principle, the quality and quantity of the data sensed, the sensor location, and the study objectives, among other issues. Ungar et al. (2005) and Augustine and Derner (2013) were the first to use a GPS collar sensor (3300LR GPS collars, Lotek Engineering, Newmarket, Ontario, Canada). Commercial accelerometer-based sensors are more readily available on the market. In this way, Roland et al. (2018) used an ear-tag sensor (Smartbow Eartag, Smartbow GmbH, Weibern, Austria), while Pavlovic et al. (2021; 2022) used a neck collar (Afimilk Silent Herdsman, NMR, Chippenham, UK). Recently, Chebli, El Otmani, Cabaraux, Keli and Chentouf (2022) and Chebli, El Otmani, Hornick, et al. (2022) combined diverse information from a GPS collar sensor (3300SL GPS collar, Lotek Wireless, Newmarket, ON, Canada) with a leg sensor with an accelerometer (IceTag, IceRobotics Ltd., Scotland, UK).

Commercial sensors based on accelerometers have been used to monitor feeding and physical activities, estimating the related parameters. Several authors (Biekker et al., 2014; Borchers, Chang, Tsai, Wadsworth, Bewley, 2016; Pereira, Heins, Endres, 2018; Zambelis, Wolfe, Vasseur, 2019) used ear-tag sensors (SensOor, CowManager) to determine rumination and eating time (feeding time). Other authors (Grinter, Campler, Costa, 2019; Werner et al., 2019) used collar sensors (MooMonitor+, Dairymaster) and Rumiwatch (Itin+Hoch GmbH, Switzerland) pressure sensor-based system (Ruuska, Kajava, Mughal, Zehner, Mononen, 2016; Steinmetz, von Soosten, Hummel, Meyer, Dänicke, 2020; Werner et al., 2018; Werner et al., 2019). Finally, rumination time was monitored with the Hitag system (Allflex), which combines an accelerometer-based collar with a sound-based device (Schirmann, von Keyserlingk, Weary, Veira, Heuwieser, 2009).

Individual feeding behaviour and feed intake for confined animals have been monitored using Insentec (Hokofarm group, the Netherlands) and Intergado (Intergado Ltd., Mina Gerais, Brazil) RFID-load cell sensor systems (Chapinal, Veira, Weary, von Keyserlingk, 2007; Chizzotti et al., 2015).

Commercial sensors have the advantage that end-users do not need to worry about technical aspects of preprocessing, feature extraction, and classification tasks. These facts simplify the data acquisition problem. However, they could be a disadvantage in research studies because of the limited flexibility in the recorded data and sensor position. Therefore, several works employed general-purpose data loggers. Vázquez-Diosdado et al. (2015) and Barker et al. (2018) used a wireless data logger that collected data from a GPS and an IMU (Omnisense Series 500 Cluster Geolocation System, Omnisense Ltd., Elsworth, UK). Fogarty, Swain, Cronin, Moraes, and Trotter (2020) and Simanungkalit et al. (2021) recorded only accelerations with a commercial data logger (Axivity AX3, Axivity Ltd, Newcastle, UK), whereas Rayas-Amor et al. (2017), Benaissa, Tuyttens, Plets, De Pessemier, et al. (2019), and Ding et al. (2022) choose to work with another commercial data logger (UA-004-64, HOBO Pendant® G Data Logger, Onset Computer Corporation) that records acceleration and tilt measurements.

## 5. Discussion

The information and communication technologies revolution will continue to have a far-reaching impact on animal farming. PLF technologies focused on monitoring animal welfare and feeding behaviour are being developed and researched. However, only a small proportion of these developments has been brought to market, and even a smaller one has been adopted by farmers. These facts arise from the complexity and multidisciplinary nature of monitoring tasks, which require balancing the needs of farmers, researchers, and animals. In the following paragraphs, we will analyse and discuss the advantages and limitations of the methodologies and algorithms.

### 5.1. Comparison of monitoring methodologies

The lack of consensus on experimental parameters (sampling time, recording period), protocols, validation strategies, and performance measures, among others, makes the comparison of monitoring methodologies difficult even for studies with the same sensing principle and goals. This situation arises because all these factors heavily depend on the experiments' aims and the final application. However, some agreements on them should be reached for each experimental goal, establishing a family of standardised experimental parameters, protocols, validation strategies, and performance measures for future works.

**Table 3**: Comparison of monitoring methodologies and their main characteristics.

| Characteristic | Movement[1] | Sound | Image[2] | Pressure |
|---|---|---|---|---|
| Allow a detailed analysis | High | Very high | Medium | High |
| Location flexibility[3] | Medium | Low | High | Very low |
| Noise robustness | Low | Very low | High | Very high |
| Wearable | Yes | Yes | No | Yes |
| Damage robustness | Very low | Very low | Very high | Very low |
| Data storage efficiency | Very high | Medium | Very low | Very high |
| Non-intrusiveness | High | High | Very high | Low |
| Device autonomy | High | Low | Very high | High |

Table 3 shows a qualitative comparison of relevant aspects of the monitoring technologies described in previous Sections. It clearly states that there is no universal monitoring technology since they have strengths and weaknesses.

---

[1] Only accelerometers, gyroscopes and magnetometers are considered in this category.
[2] Most of the characteristics for images consider them as non-wearable sensors.
[3] Typical locations of sensors and devices used for monitoring feeding behaviour are shown in Fig. 5.

Algorithms based on sound signals provide detailed information about JM and allow a precise estimation of the DMI (Galli, Cangiano, Milone, Laca, 2011; Galli et al., 2018). Raw movement signals and the associated computed features have been used to estimate the DMI using statistical and machine-learning models. Movement and pressure-based monitoring methodologies also provide temporal and frequency information regarding JMs, although less detailed than sound-based algorithms. Finally, many image-based monitoring methodologies allow the supervision of multiple animals with a single sensor, generally located far from individuals and therefore missing behavioural details at the chewing level.

Movement, sound, and pressure-based devices are wearable, allowing continuous individual supervision because they are in contact with the body of the animal. In this case, the battery life of the devices is a critical operating factor, mainly for devices that collect data at high-sampled rates (like sound) and from global satellite positioning systems. On the other hand, image-based sensors are generally not wearable, remotely sensing animals' behaviour, and have direct energy sources. Therefore, they lost details of individual feeding behaviour. Another disadvantage is the storage capacity required to save information related to high-resolution images or videos.

Sensor position is a relevant factor in algorithms based on wearable sensors. It must allow capturing behaviours without disturbing the animal and guaranteeing the sensor's integrity. Moreover, sensors must be easy to install and remove. Algorithms based on accelerometers and gyroscopes require an accurate sensor orientation to ensure the replication of the results. However, they have some flexibility in their locations, depending on the monitored behaviours. Sounds can be captured in specific positions on animals' foreheads (see Fig. 5). The location of pressure-based sensors is around the animal's mouth. Some of them can upset natural animal behaviour, disturbing the measurements. Finally, remote cameras are located in the farm infrastructure, making them the most flexible sensors in this topic.

The presence of disturbances and noises in the recorded signal deteriorates the performance of monitoring algorithms. Each sensing principle has advantages and drawbacks that must be exploited and addressed in the algorithms. In this sense, pressure-based sensors are reliable and accurate because they record the movements of the animal's jaw. They are robust against external disturbances due to noises and weather, but the sensor's parameters are time-varying, requiring continuous calibration. Image-based sensors are susceptible to changes in the scene illumination (light halos, reflections), which can be troublesome to correct or modify. Motion-based GPS sensors are unaffected by external signals when used in open fields, but the presence of buildings and solid structures degrade their performance and reliability. Motion sensors based on accelerometers and gyroscopes (including IMUs) are disturbed by vibrations and movements different from those objectives of the measurement. Another problem with these sensors is the time-varying nature of their parameters, requiring continuous calibration. Finally, sound-based sensors are susceptible to environmental noises (such as wind blowing, birds singing, and other animals) that disturb the animals' sound recordings. This problem is particularly challenging in confined environments (such as the barn) because of the sound mixing and intensity.

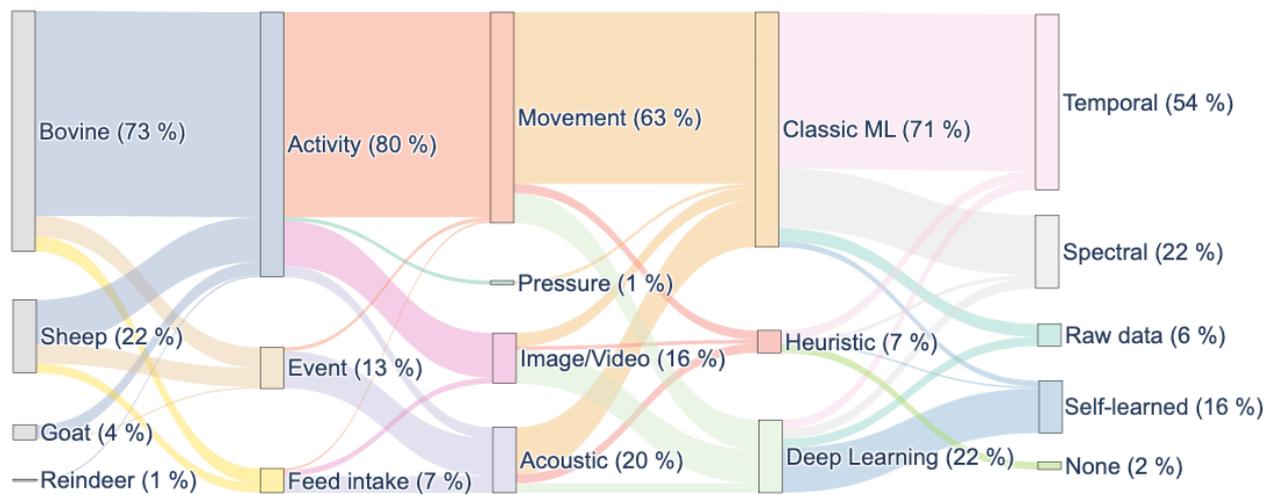

**Figure 17:** Sankey diagram showing the relationship among animal species, monitoring objectives, physical phenomena, modelling strategies, and type of features used [4].

Figure 17 shows information about the articles analysed in this work between 2005 and 2022 from different points of view using a Sankey plot, excluding those related to the validation of commercial devices. It shows the relationship between animal species, monitoring objectives, physical phenomena, classification methods, and the features used in the articles. Most of the studies were carried out in bovines (73%), followed by sheep (22%) and goats (4%). The primary objective was feeding activities recognition (80%), followed by JM event recognition (13%) and DMI estimation (7%). The physical phenomena most frequently measured were movement (63%), followed by sound (20%) and images (16%). Sound is suitable for monitoring the three objectives, especially JM recognition and feed intake estimation. Motion and image-based sensors can only monitor activities. Regarding modelling strategies, most of the published papers used classic ML (71%) and DL (22%) techniques, followed by heuristic (7%) ones. Image or video-based studies mostly used DL methods to monitor feeding behaviour. Finally, temporal features (54%) are the most commonly used type, followed by spectral (22%) and self-learned features (16%). A small percentage of studies (6%) use raw data, and the remaining do not use features (2%).

Figure 18 shows the evolution of physical phenomena (a) and computational methods (b) for monitoring ruminant feeding behaviour over time. The use of movement (Acc), sound (Mic), and image/video (Img/Vid) sensing has increased over the last two decades (Fig. 18.a). Movement sensing has expanded faster, especially since 2015. Acoustic monitoring has seen moderate adoption, providing rich behavioural information but remaining underused compared to movement. Vision-based monitoring has emerged recently, enabled by improving cameras, communications, and computer vision algorithms. Overall, the use of all three phenomena has grown, with movement leading, sound in the middle, and vision trailing but rising faster. In terms of computational methods, the use of classic ML and DL models has substantially increased over the last five years (Fig. 18.b).

---

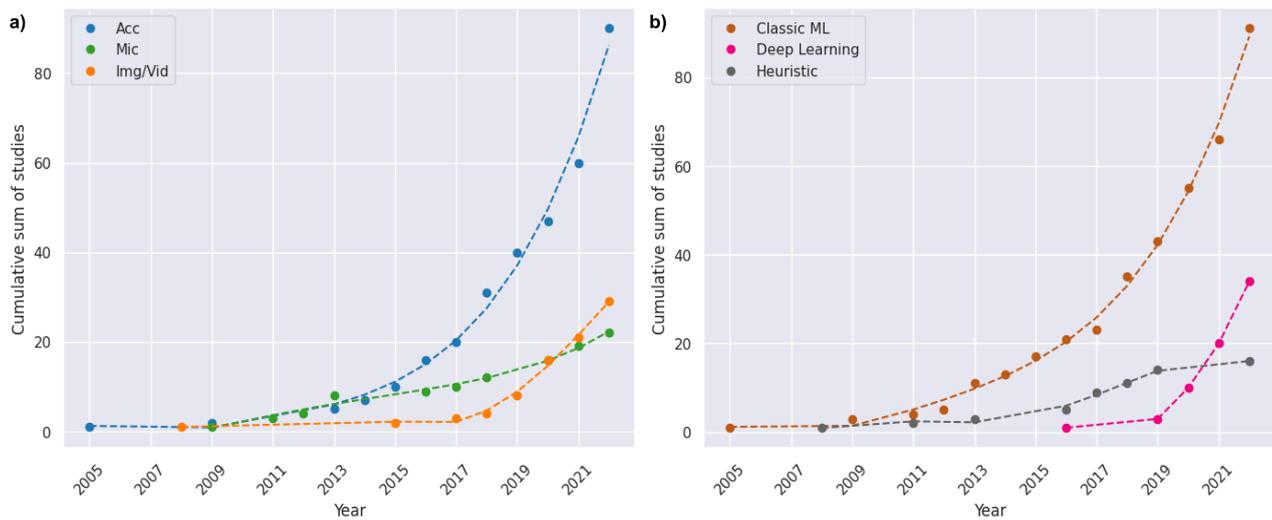

**Figure 18:** Cumulative number of articles per year describing the evolution of a) sensors and b) computational methods used to monitor feeding behaviour.

DL methods generally improve the recognition of feeding behaviour over Classic ML ones. One key advantage of DL methods is their ability to use even raw signals without any feature engineering: DL models can extract relevant features from raw data without needing manual selection or feature extraction. However, they have a higher computational load (two to three magnitude orders) than Classic ML ones. It is a significant factor in applications where real-time operation is required. However, the performance improvements may justify the additional computational resources in the case of other applications. Classic ML is a better option for portable or low-resource devices where high performance is not required. Another related issue is the number of parameters of the models. DL models typically have a large number of parameters, which increase their computational cost and memory requirements. The amount of data available for training is another issue to consider when selecting the architecture. DL models may not provide acceptable performances when the parameters-data relationship is small, as it may lead to overfitting or poor generalisation.

Models' simplicity and interpretability are other meaningful aspects to consider when choosing between DL and Classic ML methods. Classic ML methods often use white box models that are easier to interpret and understand, while DL methods use black box models that can be more difficult to interpret. It may be a relevant factor in applications where interpretability is essential (Hoxhallari, Purcell, Neubauer, 2022).

In summary, the choice between DL and Classic ML methods for monitoring feeding behaviour depends on several factors, including performance, computational cost, data availability, and interpretability. Each method has advantages and disadvantages, and the best choice will depend on the specific requirements and constraints of the application.

## 5.2. Limitations and opportunities in the field

Research groups and companies around the world are developing new techniques for monitoring animals' physical, feeding, and drinking activities. They seek changes in animal behaviours that can indicate management and disease issues or signal physiological states. This information is employed to manage and optimise farm processes by implementing better everyday herd decisions. The adoption of these technologies by end users depends on the technologies' effectiveness, validated by research groups, companies, or end users. An analogous situation occurs in academia, where other groups must be able to reproduce (validate) the results.

The proper development and assessment of algorithms require the availability of widely accepted open-access databases to develop and benchmark algorithms. A key factor for their accessibility is the cost of building. In general, databases are expensive because of the complexity and labouring efforts of recording, labelling, and curating data from experiments with many animals under different conditions. Even the availability of unlabeled databases is limited, although it could be beneficial for developing models using unsupervised or semi-supervised learning methods.

The number of animals available, recording session periods, and devices used in the experiments are fundamental factors of the experiment design. Information about the number of animals and recording periods is available in the public databases. However, the characteristics of the recording devices are often overlooked and not reported. Moreover, many methods and algorithms reported in the bibliography do not publish their source codes and parameters, which limits reproducibility.

Many limitations and issues described in the previous paragraphs arise from the lack of consensus on the experimental methodology and setups. The values of experimental parameters are selected to optimise the results, depending on the objectives. Therefore, they spread over a wide range of values. Besides, there is no clear agreement on the devices used for recording data, validation schemes, and performance measures used in the experiments. This diversity of parameters and methodology makes difficult the comparison of the algorithms, even for the same monitoring methodology.

### 5.3. Challenges and future research directions

Precision livestock farming is transforming livestock management through the integration of advanced technologies aimed at optimising resource use and enhancing animal production and welfare. To advance in this field, critical engineering requirements need to be addressed. These include the development of sophisticated sensors and real-time monitoring systems to enable accurate data collection on animal health, behaviour, and productivity. Additionally, advancements in artificial intelligence and machine learning are essential for developing algorithms capable of analysing large datasets and providing practical recommendations. The design and implementation of automation systems are necessary to facilitate precise and early interventions in livestock management. Furthermore, the integration of IoT and high-quality connectivity is crucial for ensuring efficient and secure real-time data transmission. Addressing these engineering research needs will significantly advance precision livestock farming, thereby enhancing sustainability, productivity, and animal welfare in livestock production systems. The approach to developing these management systems (platforms) is clearly interdisciplinary. The integration of knowledge from diverse fields such as animal biology (physiology, behaviour, nutrition, etc.), engineering (sensors, signal interpretation, etc.), and computer science (algorithms, artificial intelligence, etc.) is crucial for an effective and successful development.

The development and standardisation of methods to collect information that allows accurate and detailed characterisation of daily activities is a priority for future research studies. The data should be appropriate to analyse animal behaviour under different conditions, derive models for DMI prediction, detect early welfare problems, and assist in management decisions. For example, acoustic methods offer a promising approach for accurately detecting individual variations of behavioural variables relevant to herd management. Differences in grazing time, rumination time, instantaneous intake rate, and bite rate between animals or even days provide valuable diagnostic information on the limitations in feeding management. This information about the animal feeding behaviour is hard to obtain with other methods. It would be necessary to know if all methods require specific calibrations for their use in different pasture (species, phenological stages, biomasses) and animal (age, breed, frame) conditions.

Deploying feeding behaviour monitoring techniques on portable embedded systems requires further investigation and development. It is an emergent research topic known as edge artificial intelligence (EAI) that allows computations where data is collected rather than at a centralised computing facility. Because of the integration of IoT with AI, this is also known as artificial intelligence of things (AIT). Few algorithms have been implemented on resource-constrained embedded systems (Deniz et al., 2017; Arablouei et al., 2021; Yu et al., 2022). The deployment of

ML-based algorithms in low-power embedded systems comprises either the adaptation of algorithms to the available resources (hardware resources, available memory, numerical formats) or the algorithm development for the embedded system-specific data set specifications usually include using lightweight and compressed models, which results in a loss of accuracy performance (Murshed et al., 2021).

While it is argued that isolated development of EAI/AIT may be insufficient to achieve Agriculture 4.0 (Morrone, Dimauro, Gambella, Cappai, 2022), a comprehensive vision that considers different levels is necessary. In order to ensure the scalability of precision livestock solutions, it is essential to propose new systems that can distribute intelligence across several computing layers, including edge, fog and cloud (see Figure 19). This hierarchical distribution facilitates task delegation based on computational power, data privacy needs, and response time requirements. It enables real-time monitoring and analysis of large numbers of animals across vast agricultural landscapes. The framework of distributed intelligence optimises the efficiency of data processing and management, while also ensuring that farm operations remain adaptable to the evolving demands of the agricultural industry. The scalability of the systems in place is guaranteed, ensuring high levels of performance, reliability, and accuracy even as the number of animals and the complexity of farm ecosystems increase.

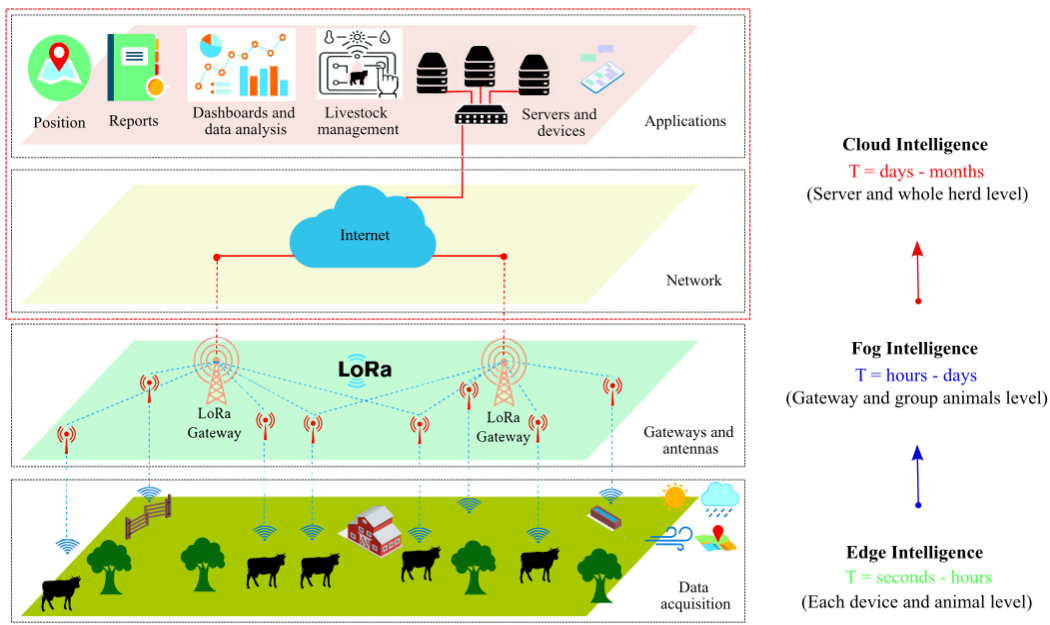

**Figure 19:** An illustrative representation of a multi-tiered intelligent ecosystem within precision livestock farming, showcasing the hierarchical flow of data through edge, fog, and cloud computing layers.

Algorithm development for embedded systems (*edge intelligence*) implies algorithm optimisation to the resources available in microcontrollers (Chelotti et al., 2016, Chelotti et al., 2018, Martinez-Rau et al., 2022). This approach has been stimulated by the availability of commercial microcontrollers with specialised hardware (floating point processor, AI accelerator/neural processor unit, encryption, security, connectivity, audio, and video interfaces). It provides algorithms with higher performance at a higher development effort, reducing the communication bandwidth and improving data security, among other features (Zhou et al. 2019). Another approach, which could be efficient in energy and performance terms, is to collect and transmit the data to be processed either on local servers (*fog intelligence*) or in the cloud (*cloud intelligence*) (Shi, Yang, Jiang, Zhang, Letaief, 2020). The optimal solution for each application will depend on the algorithm's computational cost, signal attributes, communication requirements (bandwidth, privacy, etc.), and device autonomy. However, PLF algorithms have not used this approach due to the poor communication infrastructure in rural environments.

In the search for better performance, there is a trend towards the analysis of larger volumes of data. Increasingly powerful learning methods are employed to address this challenge. Most of them employ the DL paradigm to develop classification models. The high performance obtained with these models, their ability to process unstructured data (like images or video), and the availability of efficient training methods make DL models increasingly accepted by the community. In the context of the lack of data described in the previous section, one approach to solving this problem is to generate new data of the same domain (data augmentation) or use data from different domains (data fusion). As was pointed out in previous sections, recording new data is a difficult and expensive task that research groups are not prone to carry on. Therefore, new techniques have been developed to artificially increase the size of training sets by creating modified copies of the datasets using existing data, known as *data augmentation*. These changes include the addition of noise, chunking and mixing signals portion, and using DL to generate new data points, among others.

*Domain adaptation* (Kouw & Loog, 2021) and *transfer learning* (Kleanthous, Hussain, Khan, Sneddon, Liatsis, 2022; Niu, Liu, Wang, Song, 2020) are promising techniques to address the scarcity of labelled data for training robust feeding behaviour recognition models. These methods leverage labelled data from a source domain to improve learning in a target domain with limited labelled data. For instance, models pre-trained on video/signal datasets of generic behaviours, objects, or scenes could be fine-tuned on small cattle datasets to recognise feeding behaviours. Another technique to address the lack of data is *semi-supervised learning* (Garcia, Aguilar, Toro, Pinto, Rodriguez, 2020, Yang, Song, King, Xu, 2023). It employs unlabeled data combined with a limited amount of labelled data to boost model performance. Finally, combining unlabeled and sparsely labelled cattle behavioural data could improve generalisation. Overall, these techniques may mitigate the high annotation costs and difficulty of obtaining large labelled datasets, enabling effective learning from smaller labelled datasets complemented by unlabeled or out-of-domain data (Martinez-Rau et al., 2023).

Integrating complementary data sources (*multimodal data fusion*) can lead to better recognition performances than algorithms using individual sources (Gao, Li, Chen, Zhang, 2020). This idea has been successfully employed in other research areas like human activities recognition (Nweke, Teh, Mujtaba, Al-garadi, 2019), environmental monitoring (Himeur, Rimal, Tiwary, Amira, 2022), and emotion recognition (Zhang, Yin, Chen, Nichele, 2020). However, in the topic of this review, multimodal data fusion is still a promising emergent research area since very few works have been found (Arablouei et al., 2021). The main problem to solve is the development of algorithms capable of robustly processing data from diverse domains such that they integrate information from different sources.

It was observed that precision livestock farming and particularly the monitoring of feeding behaviour in ruminants are at a turning point. This review addresses these topics but may exclude some significant studies. Given the breadth of the problem, it was difficult to cover all aspects in a single article, which is why the focus was on the areas related to the authors' expertise. It is believed that fields such as animal science or agricultural science may benefit from this research and thus expand knowledge within these domains.

## 6.   Conclusions

A review of methods and algorithms for monitoring the feeding behaviour of ruminants has been performed. Different types of sensors combined with advanced signal processing and ML techniques to assess and classify feeding activities were analysed, considering all operational aspects and features to determine their advantages and drawbacks. This evaluation includes the behavioural information provided, the sensor location on the animal, the robustness and reliability of the measurement, the device's portability and ease of use, the storage and communication requirements, the stress inflicted on the animals, and the energy efficiency of the devices.

The challenges of this research area include the requirement for additional open databases and standardised protocols to promote collaboration and secure reproducibility among researchers and developers. It will enable the comparison across studies and the validation of devices to ensure their accuracy and reliability in real-world settings. The implementation of monitoring algorithms in embedded portable devices is another relevant challenge. It is the limiting factor for the algorithms' performance since most researchers in this area do not consider this issue. Finally, algorithms based on one source of information are achieving their performance limits. Thus, there is a need for a new class of algorithms able to provide a more comprehensive understanding of ruminant feeding behaviour. They must allow the integration of different sources of information (sound, movements, images).

Precision livestock technologies must balance improving animal production efficiency with safeguarding animal health and welfare. While monitoring feeding behaviour can optimise outputs, over-focusing on enhancing productivity could compromise welfare. However, promoting humane practices may reduce short-term profits, hindering adoption unless consumer demand for sustainably produced goods increases. The goal should be to enhance both animal well-being and farm profitability, which requires a collective commitment across the supply chain to increase sustainability.

Furthermore, all these algorithms can produce valuable and timely information on animal (as well as herd) behaviour without direct human intervention, over long periods, and in locations that are difficult to access. Combined with techniques for determining environmental conditions (temperature, humidity, etc.) and pasture characteristics (forage availability and quality), they would be critical to improving the efficiency and sustainability of livestock systems. Moreover, the potential applications of these algorithms can go beyond a single farm level, including assistance in genetic and breeding evaluation, health surveillance, and animal welfare monitoring at the farm and along transport. In some countries, there are proposals to develop certification systems for livestock farming based on real-time measurements and animal behaviour as a criterion for quality labelling (Council on Animal Affairs, 2020).

Establishing these certification systems requires the development of new methodologies for data collection, processing, and integration. Collected data from different recording technologies needs to be processed and integrated into a single outcome of animal welfare, which must be easy to understand for the end-users. Finally, the integration process will require access to data from different devices and users, requiring the resolution and agreement of data ownership rights, privacy, and confidentiality issues between the parties involved. The proposed distributed intelligence model is also crucial for this integration in the roadmap towards Agriculture 4.0.

## Acknowledgements

This work has been funded by Universidad Nacional del Litoral [CAID 50620190100080LI, 50620190100151LI]; Universidad Nacional de Rosario [projects 2013-AGR216, 2016-AGR266, 80020180300053UR]; Agencia Santafesina de Ciencia, Tecnología e Innovación (ASACTEI) [project IO-2018–00082]; Consejo Nacional de Investigaciones Científicas y Técnicas (CONICET) [project 2017-PUE sinc(i)]; Knowledge Foundation [grant number NIIT 20180170]. Support program for National Universities 2023. FONDAGRO. Secretary of Agriculture, Livestock and Fisheries of the Argentine Nation. The authors would like to thank the dedication and perceptive help of Campo Experimental J. Villarino Dairy Farm staff for their assistance and support during the completion of this study.

### Contributions

**J.O.C.**: Conceptualization, Methodology, Formal analysis, Investigation, Writing - Original Draft, Visualization.

**L.S.M.R.**: Conceptualization, Data curation, Methodology, Formal analysis, Investigation, Writing - Original Draft, Visualization.

**M. F.**: Investigation, Data curation, Writing - Original Draft, Visualization, Supervision.

**L.D.V.**: Conceptualization, Methodology, Formal analysis, Investigation, Writing - Original Draft, Visualization.

**J.R.G.**: Conceptualization, Methodology, Formal analysis, Investigation, Writing - Original Draft, Visualization, Supervision, Funding acquisition.

**A.M.P.**: Methodology, Investigation, Writing - Original Draft, Visualization.

**H.L.R.**: Conceptualization, Methodology, Formal analysis, Investigation, Writing - Original Draft, Visualization, Supervision, Funding acquisition.

**L.G.**: Conceptualization, Methodology, Investigation, Writing - Original Draft, Supervision, Funding acquisition.